\documentclass[a4paper,onecolumn,11pt,unpublished]{quantumarticle}
\pdfoutput=1
\usepackage[T1]{fontenc}
\usepackage{graphicx} 
\usepackage{xfrac}
\usepackage{bm}
\usepackage{graphicx}
\usepackage{amsmath}
\usepackage{amssymb} 
\usepackage{appendix}
\usepackage{pythonhighlight}
\usepackage{enumitem}
\usepackage{braket}
\usepackage{stmaryrd}
\usepackage{multirow}
\usepackage{qcircuit}
\usepackage{soul}
\usepackage[caption = false]{subfig}
\usepackage{amsmath}
\usepackage{tabularx}
\usepackage{textgreek}
\usepackage{verbatim}
\usepackage{authblk}
\usepackage{xcolor}
\usepackage{hyperref}
\DeclareMathOperator{\iSWAP}{\mathsf{iSWAP}}
\DeclareMathOperator{\CZ}{\mathsf{CZ}}
\DeclareMathOperator{\CNOT}{\mathsf{CNOT}}
\DeclareMathOperator{\CXSWAP}{\mathsf{CXSWAP}}
\DeclareMathOperator{\SWAP}{\mathsf{SWAP}}

\begin{document}

\title{Bunny Codes: Broadening Superconducting Quantum Error Correction Capability through Advanced Control Engineering}

\author{Runshi Zhou}
\thanks{These authors contributed equally to this work.}
\affiliation{Department of Computer Science and Technology, Tsinghua University, Beijing, P.R. China}
\affiliation{TraverseQuantum Co., Ltd., Beijing, P.R. China}

\author{Xingye Yuan}
\thanks{These authors contributed equally to this work.}
\affiliation{Department of Computer Science and Technology, Tsinghua University, Beijing, P.R. China}
\affiliation{TraverseQuantum Co., Ltd., Beijing, P.R. China}

\author{Linghang Kong}
\thanks{linghang@iqubit.org}
\affiliation{Zhongguancun Laboratory, Beijing, P.R. China}

\author{Fang Zhang}
\affiliation{Zhongguancun Laboratory, Beijing, P.R. China}

\author{Kai Zhang}
\affiliation{Pengcheng Laboratory, Shenzhen, P.R. China}
\affiliation{Department of Computer Science and Technology, Tsinghua University, Beijing, P.R. China}

\author{Zhaohui Yang}
\affiliation{Department of Electronic and Computer Engineering, The Hong Kong University of Science and Technology, Hong Kong}

\author{Jianxin Chen}
\thanks{chenjianxin@tsinghua.edu.cn}
\affiliation{Department of Computer Science and Technology, Tsinghua University, Beijing, P.R. China}


\begin{abstract}
Drawing on advances in superconducting qubit control schemes that unlock enriched native gate sets at the hardware level, we systematically examine how harnessing this enlarged physical two-qubit gate pool---specifically $\CNOT$ and $\CXSWAP$---streamlines syndrome extraction for certain qLDPC codes with nonlocal stabilizers. Through an exhaustive search, we discover a set of qLDPC codes with various stabilizer weights and distances that can be implemented on the two-dimensional nearest-neighbor qubit connectivity native to superconducting hardware while achieving performance equivalent to that of the direct $\CNOT$ implementation requiring long-range interactions. We refer to those codes as Bunny codes. Across all code distances we examine, the best Bunny codes with weight-6 stabilizers in periodic boundary conditions have a code rate approximately $3\times$ that of the toric code; when converted to open boundary conditions, they retain an approximately $2\times$ code rate advantage over the rotated surface code. In circuit-level simulation, we find that some Bunny codes exhibit logical error rates an order of magnitude lower than toric codes with comparable code rates. Our results demonstrate that high-performance quantum error correction can be achieved using an expanded gate set rather than long-range couplers, thereby significantly reducing hardware complexity.
\end{abstract}

\maketitle

\section{Introduction}
Quantum computers store and process information in a Hilbert space whose dimension grows exponentially with the number of qubits, allowing them to perform certain tasks with significant speed-ups compared to classical algorithms~\cite{shor_polynomial-time_1997,grover_fast_1996}. However, the same quantum-mechanical properties that enable these advantages also make quantum computers highly susceptible to physical noise, thereby undermining the reliability of large-scale, deep quantum circuits implemented directly on physical hardware. To combat the effect of physical errors, quantum error correction codes (QECC) redundantly store the information of logical qubits into numerous physical qubits, granting it robustness against local errors.

\begin{figure}[ht]
\centering
\subfloat[\label{fig:surf_demo}]{\includegraphics[scale=0.8]{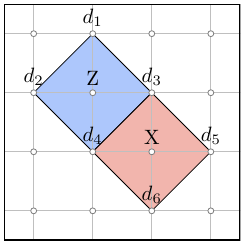}}\hspace{3mm}
\subfloat[\label{fig:cnot-only}]{\includegraphics[scale=0.72]{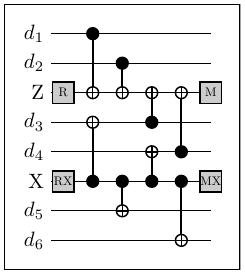}}
\caption{(a) Six data qubits (labeled $d_i$) and two ancilla qubits (labeled X and Z) in a larger surface code lattice. These qubits support two stabilizers of the surface code, $Z_1Z_2Z_3Z_4$ and $X_3X_4X_5X_6$. (b) Example of a syndrome extraction circuit for measuring these two stabilizers using only $\CNOT$ gates. Note that all $\CNOT$ gates are between adjacent qubits in the layout.}
\label{all_cnot}
\end{figure}

Superconducting quantum computing has emerged as one of the leading hardware platforms, utilizing solid-state circuits to realize qubits. This architecture offers distinct advantages, including fast and high-fidelity gate operations and high scalability. However, qubit connectivity on this platform is relatively limited because qubit couplers have to be physically installed between two qubits on the hardware to facilitate two-qubit interactions between them. Partly for this reason, previous studies have particularly favored the surface code~\cite{fowler2012surface, acharya2024quantum} because it only requires planar qubit connectivity. In a conventional surface code syndrome extraction circuit, two-qubit gates such as $\CNOT$ or $\CZ$ (which are equivalent up to single-qubit rotations) are applied between each ancilla qubit and all adjacent data qubits to extract the error syndrome associated with the corresponding Pauli stabilizer, as shown in Fig.~\ref{all_cnot}. This design works with nearest-neighbor connectivity because the surface code has local weight-4 stabilizers.

Recently, there has been growing interest in other quantum low-density parity-check (qLDPC) codes, such as hypergraph product codes (HGP codes)~\cite{Tillich_2014}, bivariate bicycle codes (BB codes)~\cite{bravyi_high-threshold_2024}, generalized bicycle codes (GB codes)~\cite{viszlai_matching_2024} and two-block group algebra codes (2BGA codes)~\cite{wang2023abeliannonabelianquantumtwoblock}. These codes offer significantly higher code rates than the surface code, but they also have stabilizers with higher weight and lower locality. With a conventional $\CNOT$-only syndrome extraction circuit design, this translates to stronger qubit connectivity requirements, such as higher degree and non-local connections.

Reconciling the connectivity requirement of high-performance quantum error correction codes with the limited connectivity on superconducting hardware has become an increasingly important problem. In conventional designs, the syndrome extraction circuit uses only $\CNOT$ gates between each ancilla qubit and an associated set of data qubits to extract the error syndrome associated with the corresponding Pauli stabilizer, as shown in Fig.~\ref{all_cnot}. Physically, $\CNOT$ is typically decomposed into native $\CZ$ gate and supplementary single-qubit rotations. To facilitate the $\CNOT$ gates, previous proposals and experimental designs~\cite{bravyi_high-threshold_2024, wang_demonstration_2025, mathews2025placingroutingnonlocalquantum} have largely pursued building quantum hardware with a multi-layer structure, in which the qubit connectivity requirements are achieved by directly installing long-range couplers.

However, there is another direction for implementing high-performance codes that has shown promise in several specific examples, yet has not been systematically examined. Namely, advanced control engineering allows superconducting platforms to calibrate and perform multiple types of two-qubit gates using their native XX+YY interactions~\cite{chen_one_2024,chen_2025_efficient,Wang_lifting_26}. In addition to the commonly used $\CNOT$ gate, the $\CXSWAP$ gate, which is logically equivalent to a $\CNOT$ gate followed by a $\SWAP$ operation, has been studied in the implementation of syndrome extraction circuits of surface code~\cite{McEwen2023relaxinghardware,eickbusch2025demonstratingdynamicsurfacecodes}. Its inherent routing capability has been leveraged to realize qLDPC codes with non-local stabilizers and improved code rates~\cite{geher_2025_directional}, as well as to substantially reduce routing overhead on connectivity-constrained hardware~\cite{yang2025qubit}.

Our work makes the assumption that the hardware supports both the $\CNOT$ gate and the $\CXSWAP$ gate as native two-qubit gates with the same gate duration. The assumption regarding an expanded native two-qubit gate set is central to our approach, as it gives qubits the flexibility to move to different positions during the syndrome extraction circuit or to remain stationary. In fact, the current common experimental practice is to calibrate and use only one type of native two-qubit gate; calibrating multiple types of two-qubit gate simultaneously might be challenging, as achieving near-optimal performance for multiple gate families simultaneously may require competing calibration objectives. Nevertheless, with rapid advances in control engineering recently~\cite{chen_2025_efficient, Wang_lifting_26}, we believe that expanding the two-qubit gate set is becoming a more practical and scalable approach compared to installing long-range couplers via multilayered hardware, which poses its own calibration challenges.

In prior work, some of the present authors adopted a top-down approach that starts with existing BB and GB codes and uses both $\CNOT$ and $\CXSWAP$ gates to implement them on hardware topologies with fewer and/or shorter long-range connections~\cite{Louvre}. In this work, we take a bottom-up approach that starts from some experimentally relevant topologies most commonly seen in superconducting hardware design, including the square grid and the hexagon grid. From these topologies, we perform an exhaustive search for efficient quantum error correction codes that can be effectively implemented on the given hardware with this expanded gate set. The same workflow naturally extends to arbitrary topologies of qubit connectivity possessing the translational symmetry described in Section~\ref{subsec:connectivity}. We refer to the collection of codes we found as ``Bunny codes'', which encompass several previously proposed codes implementable on a 2D plane, such as the La-Cross code~\cite{Louvre} and the directional code~\cite{geher_2025_directional}, as well as other codes with higher code rates for similar error-correction performances. Benefiting from the freedom of qubit routing offered by the combination of $\CNOT$ and $\CXSWAP$ gates, the best Bunny codes in periodic boundaries have code rates at least ${\sim}3$ times that of toric codes with the same code distance, with the ratio reaching $4.5$ at distance-6. In circuit-level simulation, we find that the $\llbracket40,4,5\rrbracket$ Bunny code exhibits an order of magnitude reduction in the simulated logical error rate compared to the $\llbracket 18,2,3\rrbracket $ toric code with similar code rates, under experimentally relevant noise levels. Likewise, the $\llbracket 36,8,4\rrbracket$ Bunny code has a code rate twice as high as the $\llbracket 18,2,3\rrbracket $ toric code, while reducing the logical error rate by roughly a factor of two. Some of the codes we found are listed in Table~\ref{code_parameter_pbc}. We also attempt to convert Bunny codes into open boundary codes using the method in~\cite{liang_planar_2025} so that it can be immediately implemented on the current experimental hardware. Our preliminary results still show a higher code rate of ${\sim}2$ times that of the surface code, though the improvement is less pronounced than the periodic boundary case, partly because the conversion inherently reduces the number of logical dimensions. The codes are listed in Table~\ref{code_parameter_obc}.

The rest of this work is organized as follows. In Section~\ref{sec:method} we introduce the construction of Bunny codes. In Section~\ref{sec:results} we present instances of Bunny codes with high code rate and analyze their performance via numerical simulations. We conclude this work with discussions in Section~\ref{sec:discussion}.

\emph{Note added:} At the time of finalizing this manuscript, two related preprints have appeared on arXiv~\cite{gu2026nearestneighbour,nixon2026vinecodes}, both of which presented the construction of qLDPC codes compatible with the 2D planar connectivity.  Gu \textit{et al.}~\cite{gu2026nearestneighbour} adapted the directional code~\cite{geher_2025_directional} to a grid with open boundary condition, and only utilized the $\iSWAP$ gate. Nixon \textit{et al.}~\cite{nixon2026vinecodes} investigated code construction on both periodic and open boundaries, employing a combination of the $\CZ$ gate and the $\iSWAP$ gate. In contrast, the present work provides a more general framework that permits a greater flexibility in the movement of X-ancilla and Z-ancilla. As a consequence, this approach yields code instances with more favorable parameters at a lower check weight. Furthermore, the proposed construction extends to any hardware connectivity exhibiting translational symmetry, such as the hexagonal lattice and the snub square tiling.

\section{Method}
\label{sec:method}
In this section, we outline the procedure to search for and evaluate Bunny codes for any given hardware connectivity satisfying a translational symmetry. We first define the connectivity condition using a single basic unit, assuming a translational symmetry with a period of two qubits in the horizontal and vertical directions. Then we construct the set of all possible ``actions'' available on the given hardware, where an ``action'' is a coupler-gate pair defined on the basic unit. By translational symmetry, each action corresponds to a translation-invariant layer of two-qubit gate operations acting on the entire hardware. We note how a sequence of actions corresponds to a quantum circuit satisfying the same translational symmetry. We then describe how these circuits are evaluated to determine whether they constitute valid syndrome extraction circuits that measure mutually commuting X- and Z-stabilizers. Afterward, we calculate the parameters of the resulting codes on various system sizes under both periodic and open boundary conditions.

\subsection{Hardware Connectivity}
\label{subsec:connectivity}
We begin the search by defining the connectivity of the quantum hardware. We make the assumption that the quantum hardware consists of $2l\times2m$ qubits residing at the integer grid points of a $2l\times2m$ rectangle with periodic boundary conditions. We also assume that the hardware structure possesses a discrete translational symmetry under any lattice translation by $(2x,2y)$ for arbitrary integers $x$ and $y$. Accordingly, we can divide the hardware into $l\times m$ identical square components of size $2\times2$, each containing four physical qubits. We call such a square component a ``basic unit''.

\begin{figure}[ht]
\centering
\subfloat[\label{fig:basic-unit}]{\includegraphics[scale=0.8]{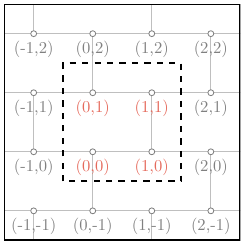}}\hspace{3mm}
\subfloat[\label{fig:hexagonal}]{\includegraphics[scale=0.8]{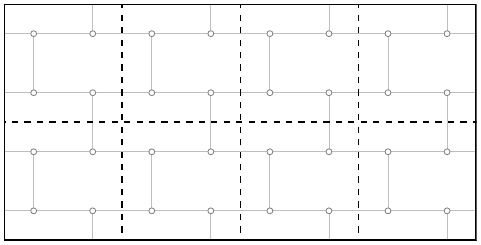}}
\caption{(a) An example of basic units on a hardware with hexagonal connectivity in the dashed box, where circles indicate qubits and solid lines between circles are couplers connecting the two qubits. The qubits are labeled by their positions in the Cartesian coordinate. (b) The hardware connectivity constructed from basic hexagonal units. Each dashed box is a basic unit that repeats on the rectangle with $l = 4$ and $m = 2$.}
\label{hex_bu}
\end{figure}

We label the qubits by their positions in a two-dimensional Cartesian coordinate system as shown in Fig.~\ref{fig:basic-unit}. Inspired by the bivariate bicycle code~\cite{bravyi_high-threshold_2024} and the generalized bicycle code~\cite{viszlai_matching_2024}, we designate two of the four qubits as X- and Z-ancilla qubits responsible for measuring the corresponding stabilizers. The other two qubits are designated as Left- and Right-data qubits. The Left- and Right-data qubits differ by their sub-lattice displacements during the syndrome extraction circuit, as we shall soon discuss. In most topologies we are interested in, the degrees of freedom associated with the initial configuration of the four qubits are redundant, as they can be absorbed in to symmetric transformations and the choice of the two-qubit gates in the first layer of the syndrome extraction circuit.

Given the translational symmetry of the hardware, describing all the couplers associated with a given basic unit, that is, couplers connected to any qubits in that basic unit, is sufficient to define the connectivity of the entire hardware, as each basic unit shares the same connectivity structure. 

For example, for a basic unit on the hardware with hexagonal connectivity, there are 9 couplers connected to its qubits, as shown in Fig.~\ref{fig:basic-unit}. We define a coupler using the notation $p_1\leftrightarrow p_2$ where $p_1$ and $p_2$ are the coordinates of the two qubits it connects. Notably, some of the 9 couplers are in fact equivalent up to translation. For example, the coupler $(-1,0)\leftrightarrow(0,0)$ is the same as the coupler $(1,0)\leftrightarrow(2,0)$ by a translation of $(2,0)$. Since our definition incorporated such translational symmetry, we regard them as equivalent representations of the same coupler. An equivalent perspective is to treat the basic unit itself as a $2\times2$ square with periodic boundaries, so that the two aforementioned couplers are indeed the same coupler. Therefore, the basic unit in Fig.~\ref{fig:basic-unit} has 6 distinct couplers, so the qubits have an average degree of 3 as expected. An example of this connectivity on a larger system is shown in Fig.~\ref{fig:hexagonal}.

\subsection{Actions}
In our bottom-up approach to searching for quantum error correction codes, we first construct practical syndrome extraction circuits based on the given connectivity, and then evaluate the error-correcting codes from the measured stabilizers.

\begin{figure}[ht]
\centering
\includegraphics[scale=0.8]{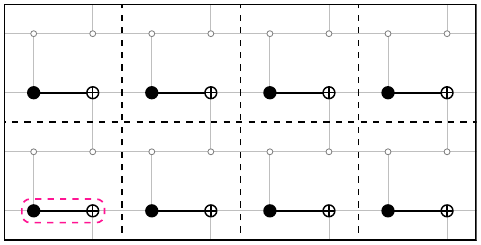}
\caption{An example of a layer of gates generated by an action of $\CNOT$ gate on coupler $(0,0)\leftrightarrow(1,0)$ in the basic unit.}
\label{ex_action_gate}
\end{figure}

In addition to the hardware structure, we further assume that in each gate layer of the syndrome extraction circuit, the same ``action'' is performed in each basic unit. That is, the same coupler in each basic unit performs the same two-qubit gate simultaneously, as shown in Fig.~\ref{ex_action_gate}. This ensures that the translational symmetry of the qubit configuration is preserved throughout the circuit. We define an ``action'' as a pair consisting of a coupler and a two-qubit gate, either a $\CNOT$ gate or a $\CXSWAP$ gate. More specifically, we assume that both $\CZ$ and $\iSWAP$ are included in the native instruction set, as they are equivalent to $\CNOT$ and $\CXSWAP$, respectively, up to single-qubit rotations. In our simulations, we assume identical gate time and fidelity for both gates, in light of the \textit{AshN} gate scheme~\cite{chen_one_2024,yang2026reconfigurable,chen_2025_efficient}. However, we note that some Bunny codes, when compared to toric codes with similar code rates, retain an advantage in logical performance even when the physical error rate is doubled. Therefore, the advantage of our scheme should persist even if the $\CXSWAP$ gate has to be implemented as two $\CNOT$ gates.

As we have previously stated, a $\CXSWAP$ gate is logically equivalent to a $\CNOT$ gate followed by a $\SWAP$ operation. While a $\CNOT$ gate directly extracts the error syndrome from a data qubit to an ancilla qubit, a $\CXSWAP$ gate not only extracts the error syndrome, but also performs qubit routing without taking additional gate time or inducing additional noise~\cite{tan2021optimal,yang2025qubit}. So, the expanded gate set essentially offers the option of performing qubit routing simultaneously with the two-qubit interaction at no additional cost. A syndrome extraction circuit with a dynamic qubit configuration constructed based on the expanded set allows qubits that are initially far apart in the original configuration to interact with each other, thus measuring non-local stabilizers. Besides, the effect of a $\SWAP$ operation can be interpreted classically as the two relevant qubits exchanging their physical positions. Thus we can always convert a syndrome extraction circuit composed of both $\CNOT$ and $\CXSWAP$ gates to a logically equivalent one with a stationary qubit configuration, in which only $\CNOT$ gates are used, but the hardware connectivity constraint is ignored. Particularly, under the SI1000 noise model~\cite{Gidney2022benchmarkingplanar}, our constructed syndrome extraction circuit should achieve comparable fault-tolerance and error-correction performance to the $\CNOT$-only implementation.

During the syndrome extraction circuit, two qubits switch their roles after a $\CXSWAP$ gate is performed between them. Since the roles of the two qubits at each end of a coupler may change, we consider an action legal only when the associated coupler connects a data qubit with an ancilla qubit. The control qubit is then determined by whether the ancilla involved is an X- or Z-ancilla.

Note that for the rest of the paper, when we refer to a qubit, we typically refer to the physical qubit that plays the role of that qubit, instead of the physical qubit at a certain fixed position on the hardware. For example, consider an X-ancilla qubit initially at position $(0,0)$. Say a $\CXSWAP$ gate is performed in the first layer of the syndrome extraction circuit using the coupler $(0,0)\leftrightarrow(1,0)$, then when we name this X-ancilla qubit at the beginning of the second layer, we refer to the physical qubit at position $(1,0)$, as it has become the X-ancilla qubit after the two qubits exchange roles. In short, we adopt the picture of dynamic qubits moving to different positions, instead of stationary qubits exchanging their information.

\subsection{Qubit Displacements and Sub-Lattices}

\begin{figure}[ht]
\centering
\subfloat{\includegraphics[scale=0.8]{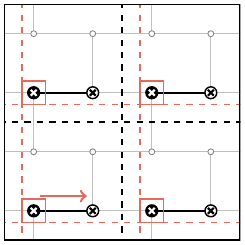}}\hspace{3mm}
\subfloat{\includegraphics[scale=0.8]{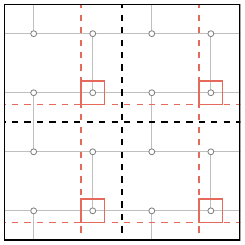}} \\
\caption{An example of the movement of a sub-lattice by a layer of gates generated by an action of $\CXSWAP$ gate on coupler $(0,0)\leftrightarrow(1,0)$ in the basic unit.}
\label{sublattice_move}
\end{figure}

The hardware qubit configuration changes during the syndrome extraction circuit as the qubits move to different positions through $\CXSWAP$ gates. As we have previously stated, since the two-qubit gates in each layer follow the translational symmetry of the hardware, the same symmetry of the qubit configuration is preserved. Consequently, qubits of a certain type (X-, Z-ancilla or Left-, Right-data qubits) always move together. For example, as the X-ancilla qubit in the basic unit moves from $(0,0)$ to $(1,0)$ by a $\CXSWAP$ gate, the action of the corresponding layer moves all X-ancilla qubits from their previous location $(2x,2y)$ to $(2x+1,2y)$ for integers $0\leq x<l$ and $0\leq y< m$. Accordingly, we keep track of the changes in the qubit configuration by recording the displacements of the four ``sub-lattices''. A ``sub-lattice'' consists of all qubits of a given type. For example, the previous example can be described as a movement of the X-ancilla sub-lattice by $(1,0)$ as shown in Fig.~\ref{sublattice_move}.

We define $D_{T}^{(i)}$ as the displacement of the sub-lattice associated with qubits of type $T$ after layer $i$ from the initial configuration at the beginning of the syndrome extraction circuit. Here $T\in\{X,Z,L,R\}$ where $X$, $Z$, $L$, and $R$ denote the X-ancilla, Z-ancilla, Left-data, and Right-data qubits, respectively. Using this definition, we can easily calculate that a qubit of type $T$ at position $a^{(i)}_T$ after layer $i$ is the qubit at position $a^{(0)}_T = a^{(i)}_T - D_T^{(i)}$ before the syndrome extraction circuit. This allows us to label every qubit by its initial position $a^{(0)}_T$ at the beginning of the syndrome extraction circuit. We can also determine the type of a given qubit $a^{(i)}$ after layer $i$ by identifying the type $T$ such that $a^{(i)} = a^{(0)}_T  + D_{T}^{(i)} + (2x,2y)$ where $a^{(0)}_T$ is the initial position of any qubit of type $T$, and $x,y\in\mathbb{Z}$ account for the translational symmetry of the qubit configuration.

Furthermore, we can use the displacements to trace back to the original positions of the interacting qubits. For example, at layer $i$ a legal action is performed using coupler $p_1\leftrightarrow p_2$, thus, without loss of generality, we assume that a $T_1$ type ancilla qubit at $a^{(i-1)} = p_1$ interacts with a $T_2$ type data qubit at $b^{(i-1)} = p_2$. We record this interaction as occurring between the ancilla qubit initially located at $a^{(0)} = a^{(i-1)} - D_{T_1}^{(i-1)}$ and the data qubit initially located at $b^{(0)} = b^{(i-1)} - D_{T_2}^{(i-1)}$ in layer $i$. By translational symmetry, this would imply that any ancilla of type $T_1$ interacts with the data qubit located at relative position $b^{(0)} - a^{(0)} = p_2 - p_1 + D_{T_1}^{(i-1)} - D_{T_2}^{(i-1)}$ in layer $i$. Therefore, we can construct a list of vectors $v_x^{(i)}$ pointing from the initial location of an X-ancilla to the initial location of the data qubit it interacts with in each layer, and similarly, this construction applies to the Z-ancilla. Note that for some values of $i$, the vector $v_x^{(i)}$ is undefined because the X-ancilla qubits do not interact with any qubits in that layer. Such a list helps to determine whether a syndrome extraction circuit is valid, as we will soon discuss in Section~\ref{subsec:search-sec}.

To keep track of the displacements of the four sub-lattices during the syndrome extraction circuit, we need to update the displacements after every layer of two-qubit gates. Following the previous example, if the gate performed is a $\CXSWAP$ gate, then we update the displacement of the $T_1$ sub-lattice by $D_{T_1}^{(i)} = D_{T_1}^{(i-1)} + p_2 - p_1$. Similarly, we update $D_{T_2}^{(i)} = D_{T_2}^{(i-1)} + p_1 - p_2$.

\subsection{Constructing Syndrome Extraction Circuits}
\label{subsec:search-sec}
\begin{figure}[ht]
\centering
\subfloat[]{\includegraphics[scale=0.6]{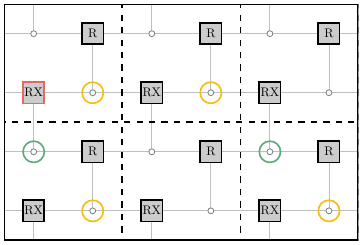}} \hspace{3mm}
\subfloat[]{\includegraphics[scale=0.6]{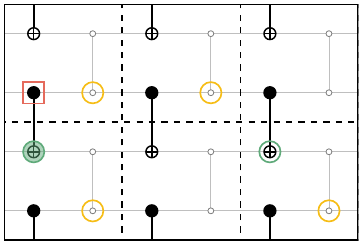}} \hspace{3mm}
\subfloat[]{\includegraphics[scale=0.6]{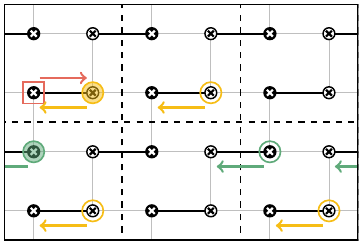}} \\
\subfloat[]{\includegraphics[scale=0.6]{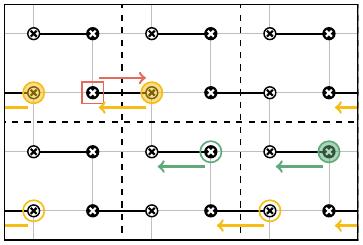}} \hspace{3mm}
\subfloat[]{\includegraphics[scale=0.6]{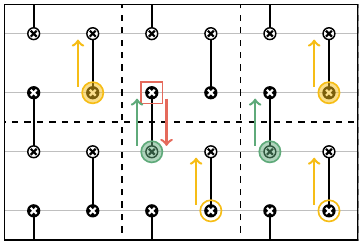}} \hspace{3mm}
\subfloat[]{\includegraphics[scale=0.6]{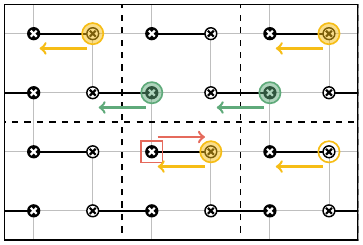}} \\
\subfloat[]{\includegraphics[scale=0.6]{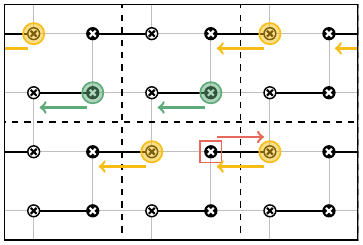}} \hspace{3mm}
\subfloat[]{\includegraphics[scale=0.6]{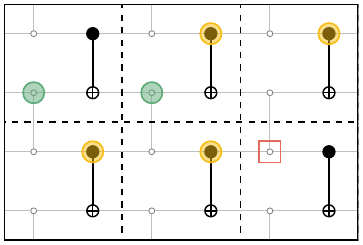}} \hspace{3mm}
\subfloat[]{\includegraphics[scale=0.6]{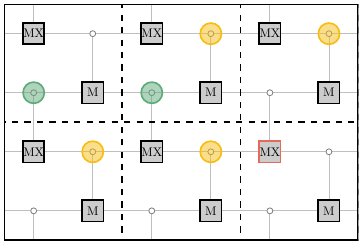}} \\
\caption{The parallelized syndrome extraction circuit constructed from the sequence of actions in Table \ref{ex_actions}, shown on a torus of size $l=3,m=2$ with hexagonal connectivity. We focus specifically on the Pauli operator measured by the X-ancilla qubit highlighted in the red box. The left (right) data qubits included in that stabilizer are highlighted in green (orange) circles, and their displacements, together with those of the corresponding sub-lattice, are tracked over the circuit.}
\label{ex_circuit}
\end{figure}

From the hardware connectivity and our assumptions of the native two-qubit gate set, we construct a set of actions that can be used to compose a syndrome extraction circuit. It is straightforward to see that a sequence of actions can be translated into a circuit. Here we walk through a simple example, with the sequence of actions defined in Table~\ref{ex_actions} and its corresponding circuit shown in Fig.~\ref{ex_circuit}.

\begin{table}[ht]
\begin{center}
\begin{tabular}{ |c|c|c|}
\hline
Index  & Coupler & Gate \\  
\hline
1 & $(0,0)\leftrightarrow(0,-1)$ & $\CNOT$\\  
\hline
2 & $(0,0)\leftrightarrow(1,0)$ & $\CXSWAP$\\  
\hline
3 & $(0,1)\leftrightarrow(-1,1)$ & $\CXSWAP$\\  
\hline
4 & $(0,0)\leftrightarrow(-1,0)$ & $\CXSWAP$\\  
\hline
5 & $(0,1)\leftrightarrow(1,1)$ & $\CXSWAP$\\  
\hline
6 & $(0,0)\leftrightarrow(0,-1)$ & $\CXSWAP$\\  
\hline
7 & $(1,0)\leftrightarrow(1,1)$ & $\CXSWAP$\\  
\hline
8 & $(0,0)\leftrightarrow(-1,0)$ & $\CXSWAP$\\  
\hline
9 & $(0,1)\leftrightarrow(1,1)$ & $\CXSWAP$\\  
\hline
10 & $(0,0)\leftrightarrow(1,0)$ & $\CXSWAP$\\  
\hline
11 & $(0,1)\leftrightarrow(-1,1)$ & $\CXSWAP$\\  
\hline
11 & $(1,0)\leftrightarrow(1,1)$ & $\CNOT$\\  
\hline
\end{tabular}
\caption{A sequence of 12 actions on hexagonal connectivity as an example.}
\label{ex_actions}
\end{center}
\end{table}

We begin the circuit by initializing all ancilla qubits in their respective bases. In this example, as well as most of the other cases considered, we choose the initial configuration so that the X-ancilla qubit is at $(0,0)$ and the Z-ancilla qubit is at $(1,1)$ in each basic unit. The Left- and Right-data qubits are placed at positions $(0,1)$ and $(1,0)$, respectively. Action 1 performs a $\CNOT$ gate using the coupler $(0,0)\leftrightarrow(0,-1)$. The action is legal since an X-ancilla qubit is at $(0,0)$ and a Right-data qubit is at $(0,-1)$ before the action is performed, which determines the qubit at $(0,0)$ to be the control qubit of the $\CNOT$ gate. Similarly, we arrange the gates for action 2, a $\CXSWAP$ gate with X-ancilla qubit as the control and Right-data qubit as the target, and action 3, a $\CXSWAP$ gate with Left-data qubit as the control and Z-ancilla as the target. Actions 2 and 3 act on disjoint sets of qubits in the basic unit, thus the two actions can be composed and performed within the same layer of two-qubit gates, as shown in layer 2 of Fig.~\ref{ex_circuit}. We note that in action 2 a $\CXSWAP$ gate is performed, displacing the sub-lattices of Right-data qubits and X-ancilla so that all X-ancilla move one lattice site to the right, and the Right-data qubits move one lattice site to the left. Likewise, in action 3 the sub-lattices of Left-data qubits and Z-ancilla are also displaced. These sub-lattice movements allow the X-ancilla labeled in red box to directly interact with the Right-data qubit labeled in orange circle by another $\CXSWAP$ gate in layer 3, action 4. Consequently, a non-local Pauli operator involving all six circled data qubits is measured by this X-ancilla, as well as all other X-ancilla qubits.

\subsection{Verifying Syndrome Extraction Circuits}
However, a globally translation-invariant circuit generated from a sequence of actions is not necessarily a valid syndrome extraction circuit that measures a set of commuting stabilizers of a quantum error correction code. To obtain such a circuit, we must ensure that the ancilla qubits are not entangled by the end of the syndrome extraction circuit, and that the stabilizers they measure all commute with each other. 

There is a simple rule to ensure both conditions. For any X-ancilla and any Z-ancilla qubit, among the set of data qubits with which they both interact during the syndrome extraction circuit, the number of data qubits that interact with the Z-ancilla before interacting with the X-ancilla must be even, and the same should hold for the number of data qubits that interact with the X-ancilla before the Z-ancilla \cite{geher2023tangling}. We can determine if a sequence of actions obeys this rule by keeping track of the qubit interactions encoded in $v_x^{(i)}$ and $v_z^{(j)}$. Relative to a fixed X-ancilla, we iterate over all $i$ and $j$ to see if $v_x^{(i)}-v_z^{(j)}$ points from the X-ancilla to a Z-ancilla, and if so, depending on whether $i>j$ or $i<j$ we increment the corresponding X-before-Z or Z-before-X count associated with that Z-ancilla. Note that $i=j$ will never happen; otherwise, the circuit would have two gates acting on the same qubit in layer $i$. In the end, we check if the number of X-before-Z and Z-before-X pairs are even for all relevant Z-ancilla, thereby verifying the aforementioned condition~\cite{zhang2026optimal}. Following the example in Fig.~\ref{ex_circuit}, an X-ancilla measures the Pauli-X operator on data qubits located at $\textbf{v}_x = ((0, -1), (1, 0), (3, 0), (4, -1), (5, -2), (7, -2),\text{Null})^T$ relative to itself in the initial configuration, while Z-ancilla measures $\textbf{v}_z = (\text{Null}, (1, 0), (3, 0), (4, -1), (5, -2), (7, -2), (8, -1))^T$. Going through the two lists, we verify, for example, $v_x^{(1)}-v_z^{(2)} = (-1,-1)$ and $v_x^{(6)} - v_z^{(7)} = (1,1)$, indicating that an X-ancilla at $(0,0)$ and a Z-ancilla at $(-1,-1)$ both interact with the data qubits at $(0,-1)$ and $(7,-2)$. From the index, we see that both of those data qubits interact with the X-ancilla before interacting with the Z-ancilla, thus the two ancilla are not entangled at the end of the syndrome extraction circuit.

Notably, the qubit configuration of the code is different before and after the syndrome extraction circuit. In most cases, the sub-lattices for Left- and Right-data qubits have different overall displacements, such that repeating the same circuit will not measure the same set of Pauli operators. A simple resolution is to perform the next round of syndrome extraction in the time-reversed order of the previous round, so that the sub-lattices are returned to the original configuration. In open boundaries, this procedure also ensures that the code moves within a finite area on the hardware. It has been proven in \cite{shaw2026optimisingquantumerrorcorrection} that the circuit-level distance of such an \textit{Alternating Circuit} is at least as large as the non-alternating version.

Our code search algorithm essentially goes through all possible sequences of actions that satisfy certain constraints, e.g., the maximum stabilizer weights of the target code. In each action, every ancilla qubit of a given type interacts with one data qubit and incorporates that data qubit into the support of a corresponding stabilizer, which is measured upon circuit completion, assuming that no two qubits interact more than once. Therefore, a code with weight-$w$ stabilizers must be generated by a sequence of $2w$ actions. Meanwhile, we also limit the maximum depth of the syndrome extraction circuit. Because some actions can be performed in parallel, a sequence of $2w$ actions can produce a circuit of any depth between $w$ and $2w$ after parallelization. We are usually more interested in the cases where the circuit depth is close to $w$.

Due to our limited computational resources, we searched for codes with stabilizer weights $w=\{4,5,6\}$ for both X- and Z-stabilizers, and for each $w$, we set the maximum depth allowed to be $w+1$ for the syndrome extraction circuit. This is inspired by~\cite{bravyi_high-threshold_2024, zhang2026optimal}, which suggests that syndrome-extraction circuits for weight-6 BB codes may require a minimum depth of 7.
During the search, we perform pruning in various aspects to mitigate the exponential growth in the number of possibilities with the length of the sequence. We evaluate whether each sequence makes a valid syndrome extraction circuit and record the valid results. Frequently, we see multiple sequences of actions leading to syndrome extraction circuits that measure equivalent stabilizers; therefore, we group the resulting sequences by the normalized shapes of the stabilizers they measure.

\subsection{Code Evaluation}

After the search, we evaluate the quantum error correction codes generated by the measured stabilizers on lattices of various sizes. We parameterize the lattice size by $l$ and $m$, which are the length and width of the rectangular torus measured in the number of basic units. Different values of $l$ and $m$ can lead to codes with different parameters. For example, we found that the syndrome extraction circuit in Fig.~\ref{ex_circuit} produces a $\llbracket12,4,3\rrbracket$ code with $l=3$ and $m=2$. Similarly, it produces a $\llbracket18,6,3\rrbracket$ code with $l=3$, $m=3$ and a $\llbracket 30,6,4\rrbracket$ code with $l=5$, $m=3$. Note that certain codes might have better parameters on more generalized parallelograms with periodic boundaries, such as the rotated toric code~\cite{Kovalev_2012}. Here, we consider only rectangles because they are easier to adapt to open boundaries. We use a ring-theoretic algorithm~\cite{chen2025generalized} for the logical dimensions, and an integer-programming-based algorithm~\cite{bravyi_high-threshold_2024, landahl2011faulttolerantquantumcomputingcolor} to calculate the code distance of codes with various values of $l$ and $m$. Since the computational cost for distance calculation could scale exponentially with the size and distance of the code, we do not perform a complete search over the entire search space. With a given pair of stabilizers, we terminate the code distance calculation for $l$ and $m$ if the number of data qubits exceeds $n_{max}$, or a strictly smaller code (with both $l$ and $m$ less than the current value) has been found with distance no less than $d_{max}$.

Since the syndrome extraction circuits are specified by the corresponding sequence of actions, we also calculate the circuit-level distance for those circuits using \textit{pysat}~\cite{imms-sat18,itk-sat24}, as a predictor of their performance under circuit-level simulation. We also simulate the logical error rates of some of the codes we found under the SI1000 noise model~\cite{Gidney2022benchmarkingplanar} using \textit{Stim}~\cite{gidney_stim_2021} with the Relay-BP decoder~\cite{muller2025improvedbeliefpropagationsufficient}, during which the number of shots is chosen adaptively to maintain a standard deviation of approximately $5\%$ of the logical error rate.

\subsection{Towards Open Boundary}
We also convert the resulting quantum error correction codes from periodic to open boundaries using existing protocols~\cite{liang_planar_2025, eberhardt_pruning_2024}, so that they can be implemented more easily on current hardware. In these protocols, converting a qLDPC code from periodic to open boundary conditions halves the number of logical qubits while preserving the physical qubit count and code distance, effectively halving the code rate. The stabilizer shapes of the open boundary code will either be the same, or become a subset of the shape of the original stabilizers, and thus can be measured with the same syndrome extraction circuit with the help of padding qubits. 

\begin{figure}[ht]
\centering
\includegraphics[scale=1.0]{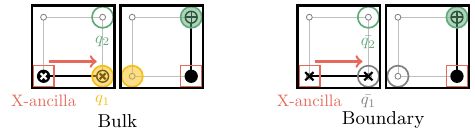}
\caption{A demonstration of using padding qubits to facilitate routing at the boundary of an open-boundary code. The X-ancilla (red box) performs a $\SWAP$ gate with $\bar{q_1}$ (gray circle) before it interacts with $\bar{q_2}$ (green circle), whereas in the bulk region, the X-ancilla (red box) performs a $\CXSWAP$ gate with $q_1$ (yellow circle) before it interacts with $q_2$ (green circle).}
\label{padding_demo}
\end{figure}

For example, as shown in Fig.~\ref{padding_demo}, suppose that in the periodic boundary code, an X-ancilla performs a $\CXSWAP$ with $q_1$, followed by a $\CNOT$ gate with $q_2$. Both qubits belong to the support of the stabilizer it measures. However, in the open boundary code, this X-ancilla is located near the code boundary, and the new stabilizer it measures no longer contains $q_1$. Since the bulk region of the hardware still satisfies the assumption of uniform connectivity with translational symmetry, a qubit $\bar{q_1}$ should exist in the position of $q_1$, although it does not serve as a data qubit of the code. We use the qubit $\bar{q_1}$ as a padding qubit and perform a $\SWAP$ gate between the X-ancilla and the padding qubit. This preserves a uniform sub-lattice motion of all X-ancilla qubits, allowing the boundary X-ancilla to interact with the qubit $\bar{q_2}$ corresponding to $q_2$ in the next step. The $\SWAP$ gate can be converted to a $\CXSWAP$ gate by setting the padding qubit in the state $\ket{0}$ or $\ket{+}$.

In addition, open boundary codes from the aforementioned conversion are typically redundant, so that the number of physical qubits can often be reduced without changing either the code distance or the number of logical qubits, requiring only minor modification to the stabilizers of the code. For example, the rotated surface code with a code rate of $\sfrac{1}{d^2}$ is obtained by transforming the unrotated surface code with a code rate of ${\sim}\sfrac{1}{2d^2}$, so that the code rate remains the same as that of the toric code in the periodic boundary condition. To meet our specific requirements, we combine two reduction algorithms. The shallow reduction is resource efficient and deterministic. It essentially removes the apparently redundant parts of the code:
\begin{enumerate}
    \item Remove all weight-1 stabilizers and their corresponding ancilla.
    \item Remove all data qubits that participate exclusively in X-stabilizers or Z-stabilizers. If any data qubit is removed during this step, return to step 1.
\end{enumerate}
To further enhance the code rate, we perform a deep reduction, which is a greedy algorithm inspired by \cite{liang_planar_2025}:
\begin{enumerate}
    \item Label all stabilizers as ``unmarked''.
    \item Tentatively remove an ``unmarked'' stabilizer, giving priority to stabilizers with smaller weights.
    \item Perform the shallow reduction.
    \item Calculate the code distance and the number of logical dimensions. If both are maintained, keep the change and return to step 1. If not, revert the change and mark this stabilizer as ``marked''. Return to step 2.
    \item Stop when all stabilizers are ``marked''.
\end{enumerate}
It is straightforward to see that, since both the shallow and deep reduction algorithms modify the code only by removing stabilizers or data qubits, the support of every remaining stabilizer is contained within that of a corresponding stabilizer in the periodic-boundary code, and thus can be measured using the same syndrome extraction circuit with the help of padding qubits. However, the deep reduction algorithm imposes a significant computational overhead, as it requires an iterative evaluation of the distance of the code. Furthermore, the algorithm is nondeterministic, as multiple stabilizers could all be legally removed individually but not simultaneously at a certain step. So, occasionally, we have to make many attempts at a code before reaching a near-optimal result. Consequently, we only apply the reduction procedure to the codes that appear the most promising, rather than exhaustively reducing every code found in the search.

\section{Results}
\label{sec:results}
\begin{figure}[ht]
\centering
\subfloat[]{\includegraphics[scale=1]{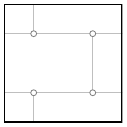}}\hspace{3mm}
\subfloat[]{\includegraphics[scale=1]{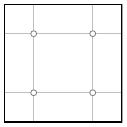}}\hspace{3mm} 
\subfloat[]{\includegraphics[scale=1]{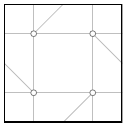}} \\
\caption{Diagrams for some qubit connectivities used in this paper. (a) Hexagons. (b) Squares. (c) Snub Squares.}
\label{bus}
\end{figure}

During our search, we have used several different topologies for qubit connectivity on quantum hardware, including hexagons, squares, and snub squares, as shown in Fig.~\ref{bus}.

The time overhead for searching different topologies depends on their degree and their symmetry properties, but as the length of the action sequence increases, an exponential growth seems unavoidable.

\subsection{Periodic Boundaries}
For periodic boundaries, we scan different values of $l$ and $m$ ranging from 1 to 20, with the number of data qubits no more than $n_{max} = 200$, and an upper bound of distance is set to $d_{max} = 8$. Our results are presented in Table~\ref{PBC_code_distance}.

\begin{table}[ht]
\begin{center}
\begin{tabular}{ |c|c|c|c|c|c|}
\hline
 \multicolumn{6}{|c|}{Weight-4 Stabilizers}  \\ 
 \hline
Distance  & 3 & 4 & 5 & 6 & 7\\  
\hline
Hexagon  & $\sfrac{1}{6}$ & $\sfrac{1}{8}$ & $\sfrac{1}{15}$ & $\sfrac{1}{18}$ & $\sfrac{1}{28}$\\  
\hline
Square  & $\sfrac{1}{5}$ & $\sfrac{1}{8}$ & $\sfrac{1}{14}$ & $\sfrac{1}{18}$ & $\sfrac{1}{28}$\\  
\hline
Snub Square  & $\sfrac{1}{5}$ & $\sfrac{1}{8}$ & $\sfrac{1}{14}$ & $\sfrac{1}{18}$ & $\sfrac{1}{28}$\\   
\hline
\hline
 \multicolumn{6}{|c|}{Weight-5 Stabilizers}  \\ 
 \hline
Distance  & 3 & 4 & 5 & 6 & 7\\  
\hline
Hexagon  & $\sfrac{2}{9}$ & $\sfrac{1}{9}$ & $\sfrac{2}{27}$ & $\sfrac{1}{15}$ & $\sfrac{2}{45}$\\  
\hline
Square  & $\sfrac{2}{9}$ & $\sfrac{1}{6}$ & $\sfrac{2}{15}$ & $\sfrac{1}{9}$ & $\sfrac{1}{21}$\\  
\hline
Snub Square  & $\sfrac{2}{9}$ & $\sfrac{1}{6}$ & $\sfrac{2}{15}$ & $\sfrac{1}{9}$ & $\sfrac{1}{21}$\\ 
\hline
\hline
 \multicolumn{6}{|c|}{Weight-6 Stabilizer}  \\ 
 \hline
Distance  & 3 & 4 & 5 & 6 & 7\\  
\hline
Hexagon  & $\sfrac{1}{3}$ & $\sfrac{1}{4}$ & $\sfrac{1}{10}$ & $\sfrac{1}{8}$ & $\sfrac{1}{20}$\\  
\hline
Square  & $\sfrac{1}{3}$ & $\sfrac{1}{4}$ & $\sfrac{1}{7}$ & $\sfrac{4}{27}$ & $\sfrac{1}{14}$\\  
\hline
\end{tabular}
\caption{Highest code rates achieved by Bunny codes for each code distance on different qubit connectivities under periodic boundary conditions.}
\label{PBC_code_distance}
\end{center}
\end{table}

Our search covers many previous results, including a toric code version of the surface code on hexagonal topology~\cite{McEwen2023relaxinghardware}, La-Cross code on grid topology~\cite{Louvre} and the directional code~\cite{geher_2025_directional}. The directional code with weight-5 stabilizers has code rates of $\frac{1}{9}$ and $\frac{1}{18}$ for code distances 4 and 6, respectively, in periodic boundaries, which match or are slightly below the best Bunny code we found. Meanwhile, our results significantly outperform their best codes with weight-6 stabilizers at code distances 4 and 6, which have code rates of $\frac{2}{21}$ and $\frac{1}{24}$, respectively.

There are a few things to note. First, in Fig.~\ref{bus}, the ``snub square'' connectivity has two extra couplers in addition to those of the ``squares'' connectivity and thus has a degree of 5. It has more possible actions, but less room for pruning because it possesses fewer symmetries than the ``squares'' connectivity. Consequently, searching for codes on this connectivity with weight-6 stabilizers was too computationally demanding for the resources available to us. Nevertheless, it is interesting that the additional couplers do not lead to higher-rate quantum error correction codes with weight-4 and weight-5 stabilizers. This observation suggests that naively adding more couplers to the qubit connectivity may not significantly improve the achievable codes rates. For comparison, the famous $\llbracket 72,12,6\rrbracket$ BB code has a code rate of $\frac{1}{6}$, which is close to our best result of $\frac{4}{27}$ but requires a significant amount of long-range connections.

Second, among the codes we found, most have X-stabilizer shapes that are the same as the Z-stabilizer shapes rotated by 180 degrees. Although allowed by our construction, codes with X- and Z-stabilizers of different shapes typically have significantly poorer parameters. This pattern persists across all stabilizer weights and connectivity that we examined.

As we have previously stated, we have also calculated the circuit-level distances of some of the most competitive codes. The underlying algorithm essentially works by solving the detector error model of the syndrome extraction circuit as a boolean satisfiability problem, which is computationally demanding. Thus, we only calculate the circuit-level distances for codes with code distances ranging from 3 to 5. 

\begin{figure}[ht]
\centering
\subfloat[]{\includegraphics[width = 0.45 \columnwidth]{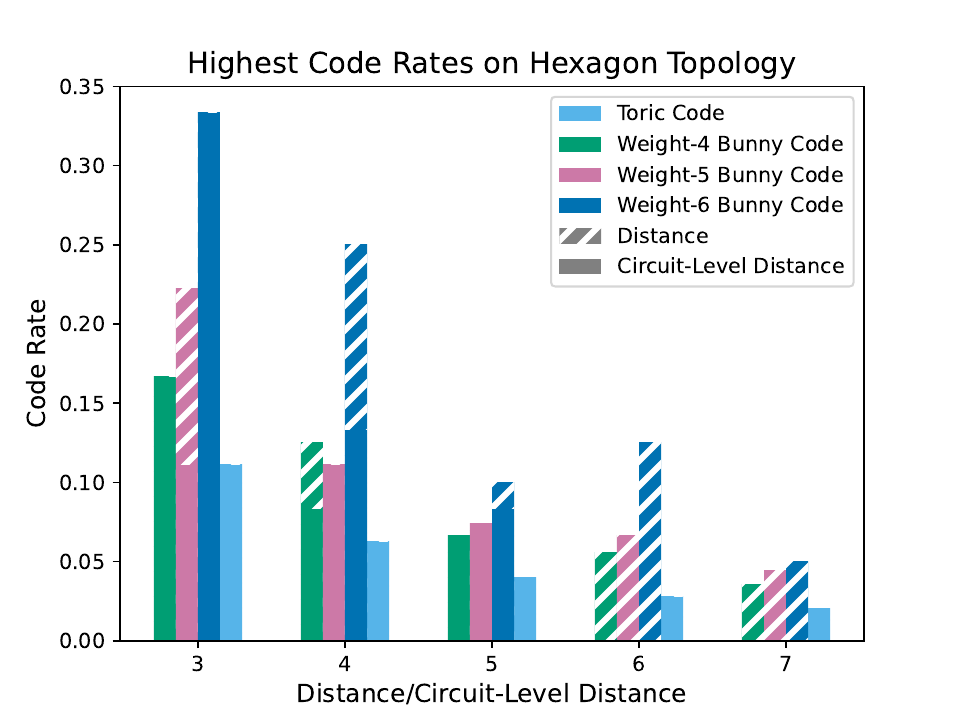}}\hspace{3mm}
\subfloat[]{\includegraphics[width = 0.45 \columnwidth]{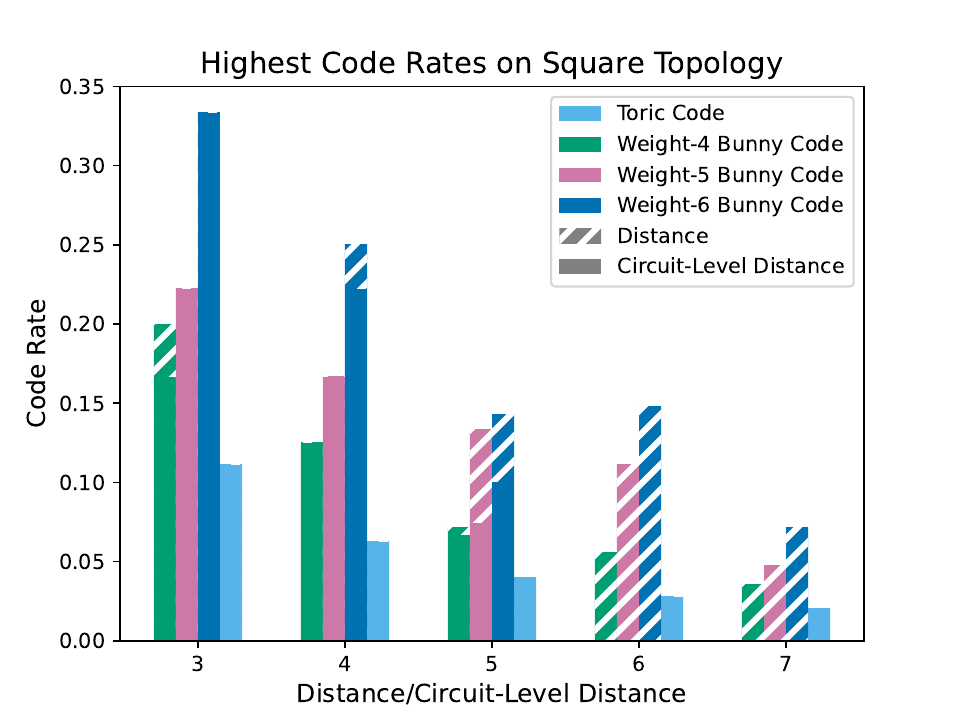}}\\
\caption{Highest code rates achieved for quantum error correction codes found with different connectivities topologies under periodic boundary conditions. The hatched region correspond to the code distance, while the solid part correspond to the circuit-level distance. Circuit-level distances are only calculated for codes with code distances up to 5. (a) Hexagons. (b) Squares.}
\label{d_n_cld}
\end{figure}

Our results are shown in Fig.~\ref{d_n_cld}. The hatched parts of the diagrams are results from Table \ref{PBC_code_distance}, while the solid parts account for the circuit-level distances. It exhibits the highest code rate achievable with the given stabilizer weights and connectivity at each code distance and circuit-level distance.  Notably, in many cases, codes with the highest code rate do not have a syndrome extraction circuit that is implementable with the given connectivity that saturates the code distance.

\begin{figure}[ht]
\centering
\includegraphics[width = 0.45 \columnwidth]{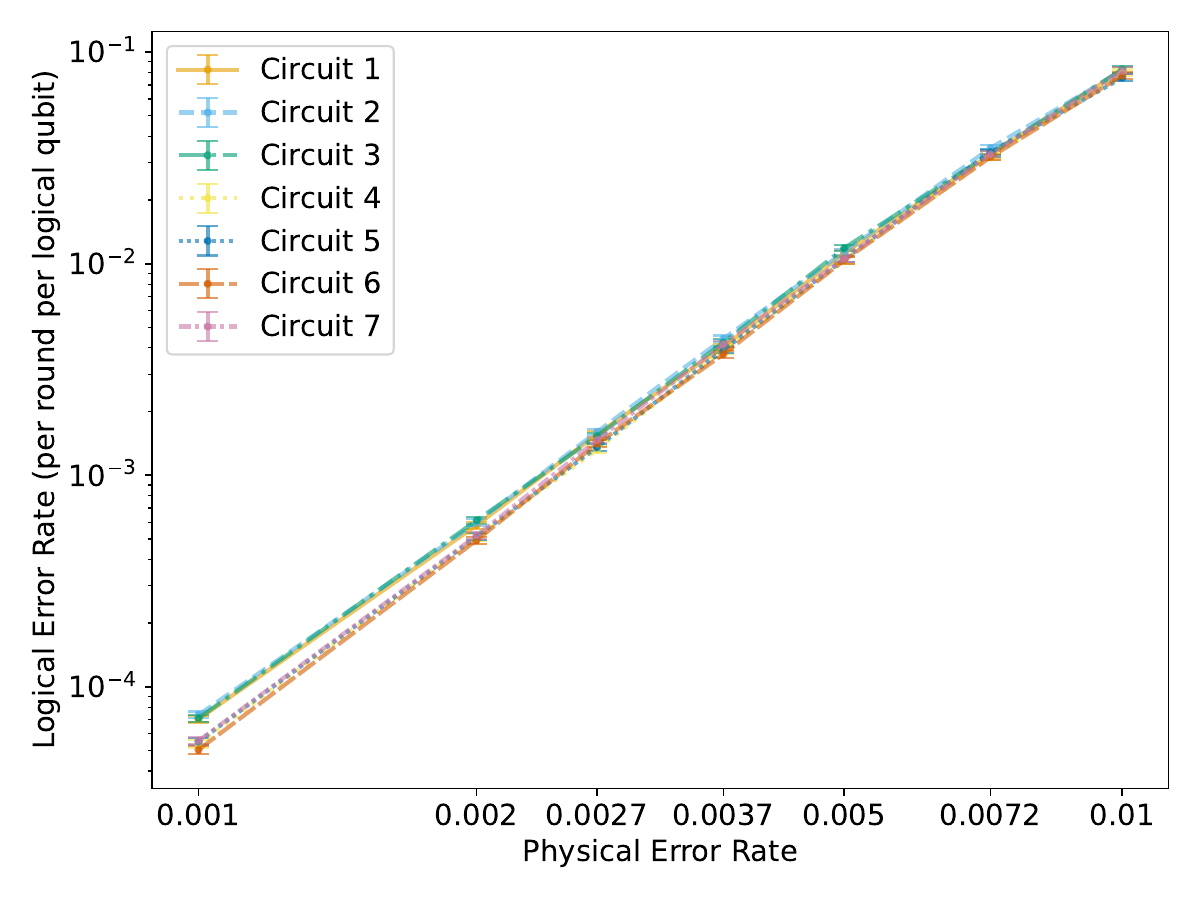}
\caption{The logical error rate per round per logical qubit of the $\llbracket 30,2,5\rrbracket$ code. Circuits 1--3 have a circuit-level distance of 3, and circuits 4--7 have a circuit-level distance of 5.}
\label{cld_ler_demo}
\end{figure}

However, through circuit-level simulation, we found that the significance of circuit-level distance is sometimes less clear than expected. For example, as shown in Fig.~\ref{cld_ler_demo}, we simulated the logical error rate of 7 different circuits that measure the equivalent stabilizers of a $\llbracket 30,2,5\rrbracket$ code. Among the 7 circuits, circuits 1--3 have a circuit-level distance of 3, and circuits 4--7 have a circuit-level distance of 5. However, when the physical error rate of two-qubit gates is $10^{-3}$, the logical error rate of circuits 1--3 is only higher than that of the other circuits by ${\sim}50\%$, a relatively modest difference.

\begin{figure}[ht]
\centering
\subfloat[]{\includegraphics[width = 0.33 \columnwidth]{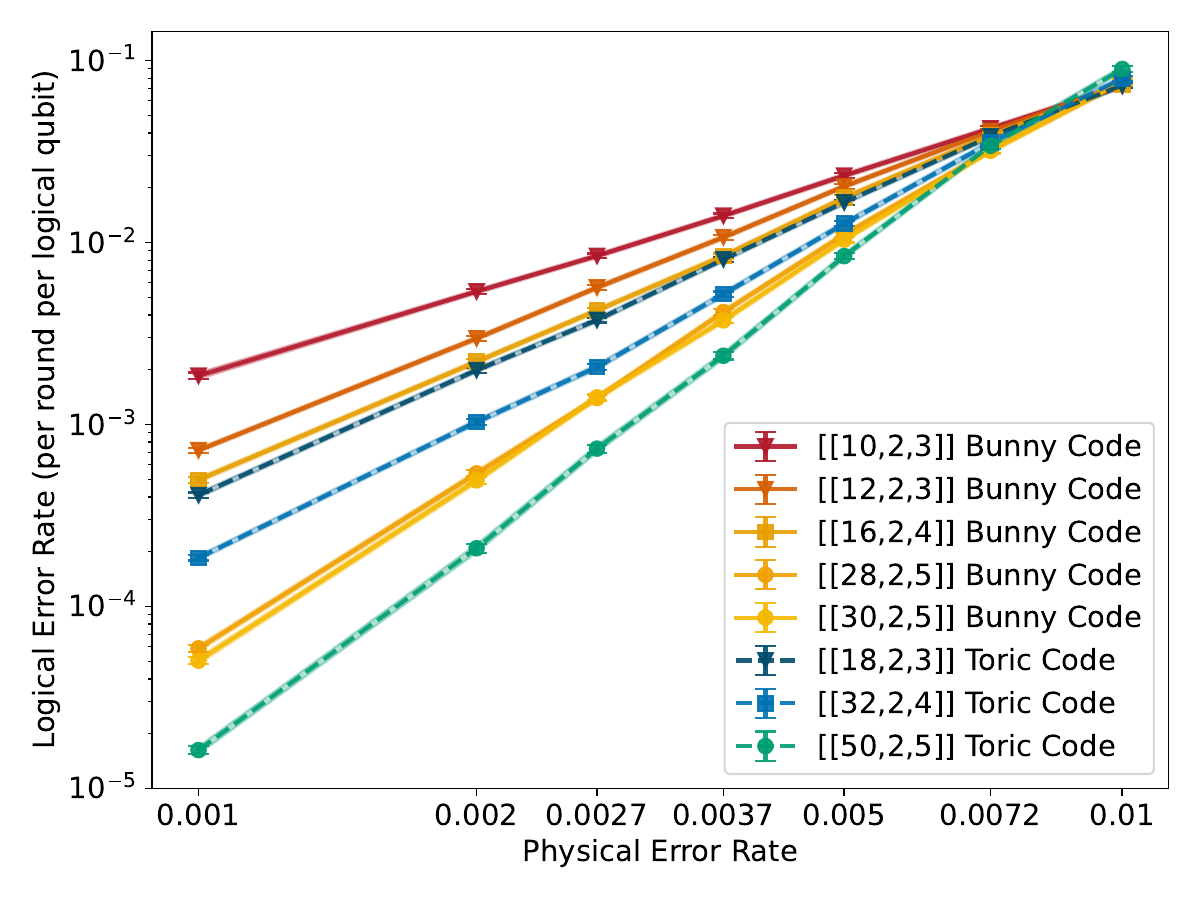}}
\subfloat[]{\includegraphics[width = 0.33 \columnwidth]{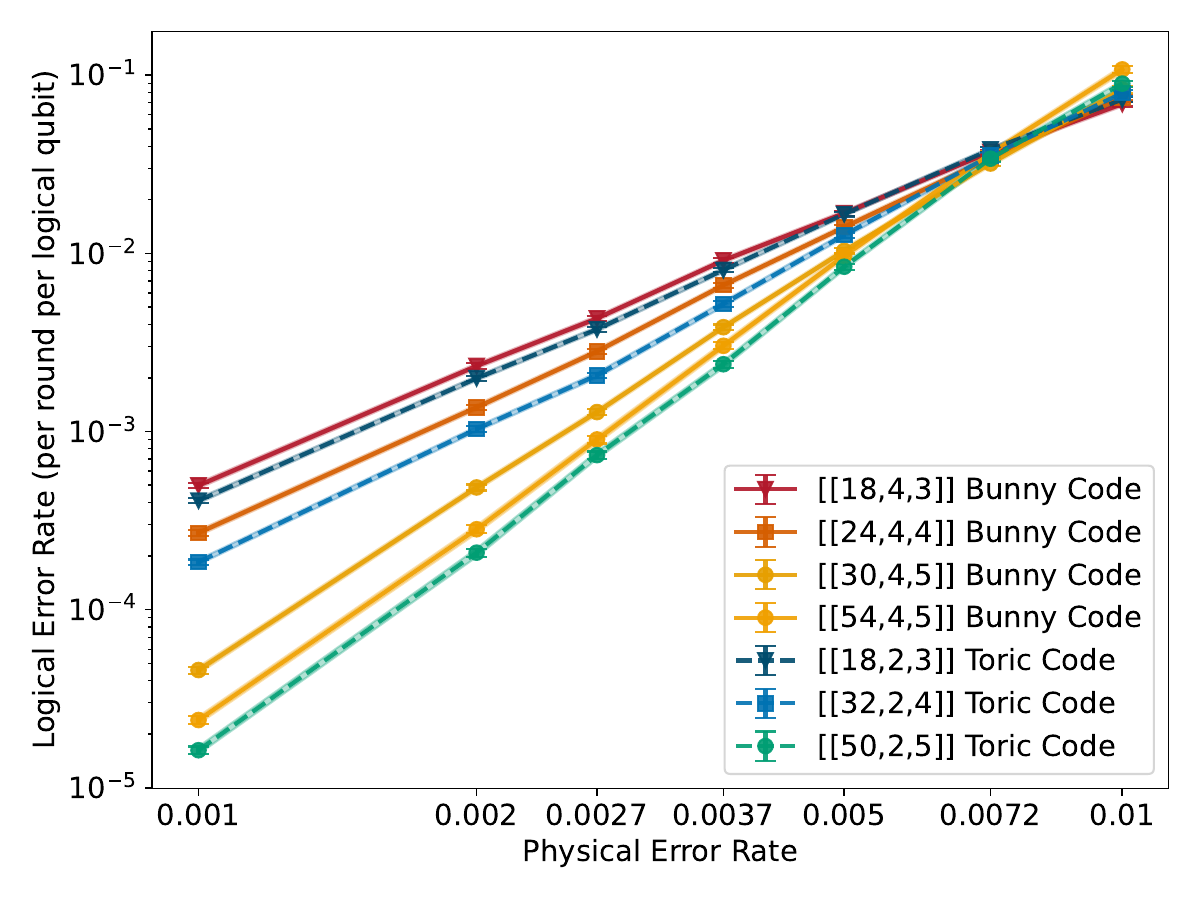}}
\subfloat[]{\includegraphics[width = 0.33 \columnwidth]{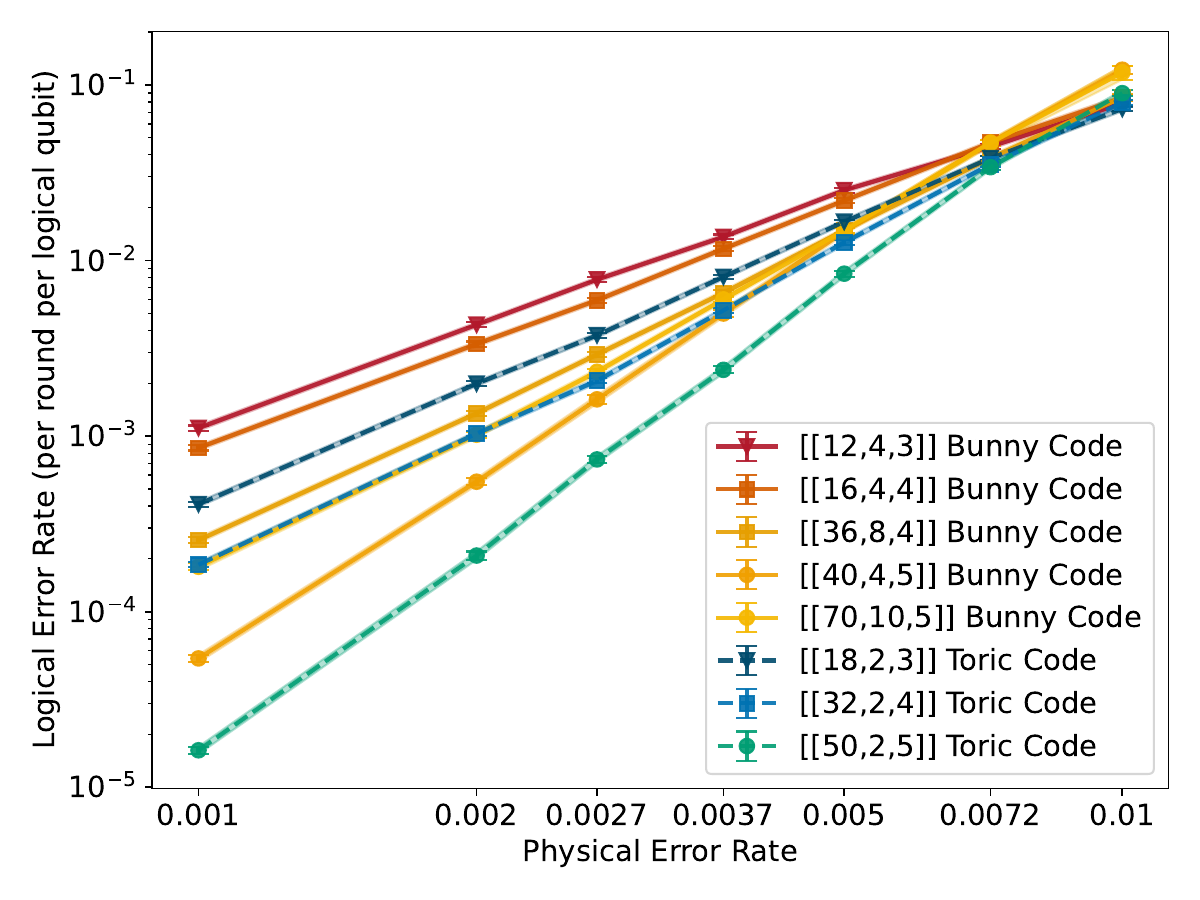}}\\
\caption{The logical error rate per round per logical qubit for some representative Bunny codes under periodic boundary conditions. (a) Weight-4 stabilizers. (b) Weight-5 stabilizers. (c) Weight-6 stabilizers.}
\label{ler_pbc}
\end{figure}

We have performed circuit-level simulations to estimate the logical error rates per round per logical qubit of several representative Bunny codes, and compared them to those of the toric code. The results are shown in Fig.~\ref{ler_pbc}. The $\llbracket 36,8,4\rrbracket$ Bunny code with weight-6 stabilizers, for example, has a code rate twice as high as the $\llbracket 18,2,3\rrbracket$ toric code, while the logical error rate at the physical error rate of $10^{-3}$ is only approximately $60\%$ of that of the toric code. Similarly, the $\llbracket 40,4,5\rrbracket$ Bunny code has a code rate comparable to that of the aforementioned toric code, while exhibiting a logical error rate ${\sim}7.5$ times lower and a different scaling trend in the low-noise regime.

\begin{figure}[ht]
\centering
\includegraphics[width = 0.6 \columnwidth]{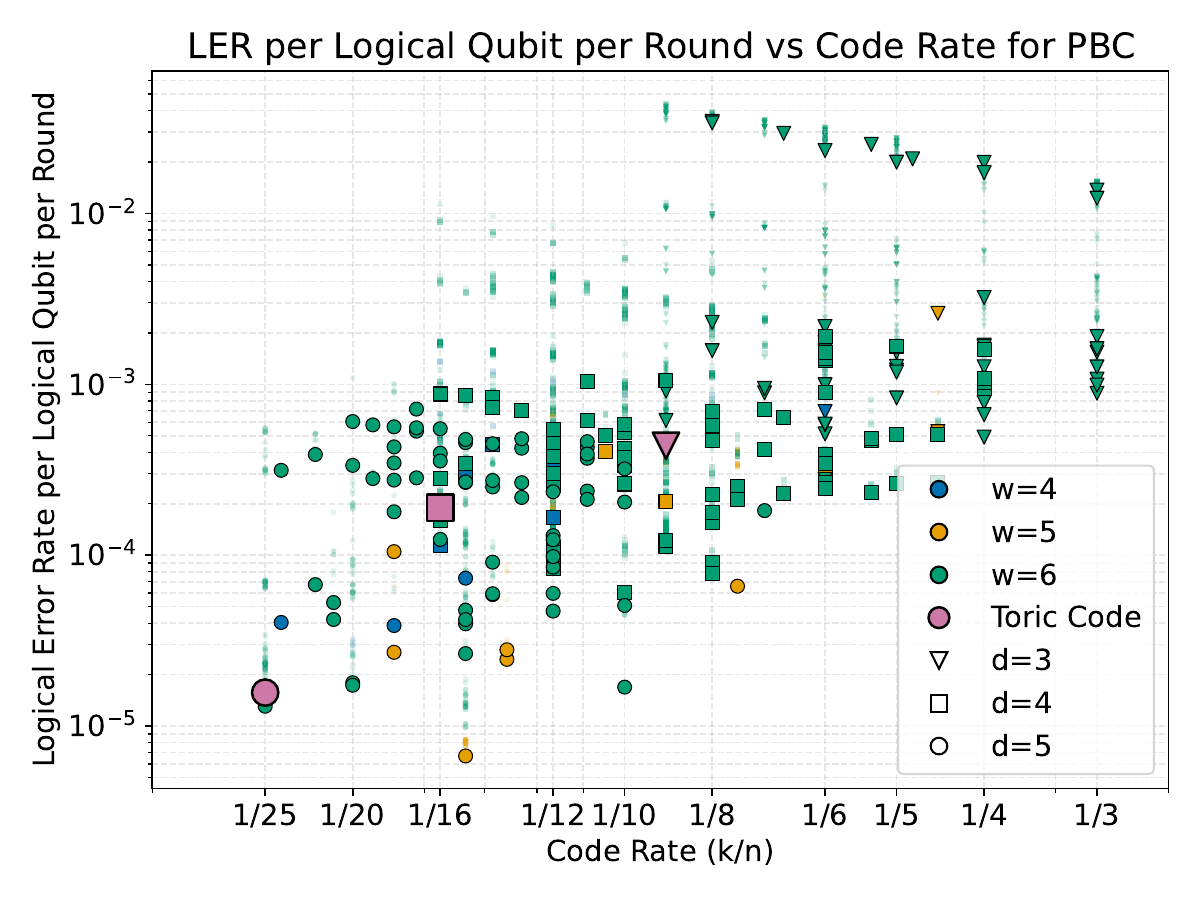}
\caption{The logical error rates per round per logical qubit of periodic-boundary Bunny codes and toric codes at a physical noise level of 0.001. For clarity of presentation, codes are grouped by their parameters $(l,m,n,k,d)$ and with in each group only the code with the best logical performance is highlighted while the other codes are faded out. The colors of the markers correspond to the weights of their stabilizers, while the shapes correspond to their code distances. The performance of toric codes is highlighted in pink for reference.}
\label{pbc_square_scatter}
\end{figure}

We also performed a larger-scale simulation surveying all the periodic-boundary Bunny codes implementable on the square topology with advantageous code rates ($\frac{kd^2}{n}\geq1$) and reasonable code sizes ($n\leq 3d^2$). The circuit-level simulation was performed at a physical error rate of 0.001, and the logical error rates per logical qubit per round of these codes are shown in Fig.~\ref{pbc_square_scatter}.

Although our framework does not impose a symmetry between the shapes of X- and Z-stabilizers, we find that codes with highest code rates all have such a symmetry, so they can be described in the language of BB and GB codes. The parameters and generating polynomials of some of the codes we found are listed in Table \ref{code_poly}. Note that for each set of $n,k,d$ we usually have more than one set of stabilizers, each with multiple syndrome extraction circuits that can be implemented on the connectivity given, here we only list a small portion of them for demonstration.

\begingroup
\renewcommand{\arraystretch}{0.7}
\begin{table}[ht]
\small
\begin{center}
\caption{The parameters and generating polynomials of some of the periodic boundary codes discussed in this paper.}
\label{code_parameter_pbc}
\begin{tabular}{ |c|c|c|c|c|c|c| } 
\hline
$\llbracket n,k,d\rrbracket$&$l$&$m$& $kd^2/n$&  $A$ & $B$ & Connectivity\\ 
\hline
\hline
$\llbracket12,2,3\rrbracket$& 3&2& 1.33 & \multirow{4}{*}{$1+x$} & \multirow{4}{*}{$x+y$} & \multirow{4}{*}{Hexagon}\\ 
\cline{0-3}
$\llbracket16,2,4\rrbracket$& 4&2& 2 &  &  & \\ 
\cline{0-3}
$\llbracket24,2,4\rrbracket$& 4&3& 1.33 &  &  & \\ 
\cline{0-3}
$\llbracket30,2,5\rrbracket$& 5&3& 1.66 &  &  & \\ 
\hline
$\llbracket10,2,3\rrbracket$& 1&5& 1.8 &  \multirow{2}{*}{$x+y^2$} & \multirow{2}{*}{$y+y^2$} & \multirow{2}{*}{Square} \\ 
\cline{0-3}
$\llbracket28,2,5\rrbracket$& 2&7& 1.79 &  &  & \\ 
\hline
\hline 
$\llbracket18,4,3\rrbracket$& 3&3& 4 & \multirow{2}{*}{$x+xy+x^2y$} & \multirow{2}{*}{$xy+y^2$} & \multirow{4}{*}{Hexagon}\\  
\cline{0-3}
$\llbracket36,4,4\rrbracket$& 3&6& 1.78 &  &  & \\  
\cline{0-5}
$\llbracket36,4,4\rrbracket$& 6&3& 1.78 & \multirow{2}{*}{$1 + x + x^2$} & \multirow{2}{*}{$x+y$} & \\  
\cline{0-3}
$\llbracket54,4,5\rrbracket$& 9&3& 1.85 &  &  & \\  
\hline
$\llbracket18,4,3\rrbracket$& 3&3&  4 & \multirow{2}{*}{$1 + xy + x^2y$} &  \multirow{2}{*}{$x + xy$} & \multirow{3}{*}{Square}\\  
\cline{0-3}
$\llbracket24,4,4\rrbracket$& 3&4& 2.66 &  & &\\  
\cline{0-5}
$\llbracket30,4,5\rrbracket$& 3&5& 3.33 & $1 + x + x^2y^2$ & $xy + xy^2$ & \\ 
\hline
\hline 
$\llbracket12,4,3\rrbracket$& 3&2& 3 & $x+y+xy+x^2$ & $xy+x^2y$ & \multirow{3}{*}{Hexagon}\\
\cline{0-5}
$\llbracket18,6,3\rrbracket$& 3&3& 3 & $ y + xy + x^2 + x^3$ & $xy + x^3y$ & \\
\cline{0-5}
$\llbracket16,4,4\rrbracket$& 4&2& 4 & $xy + x^2y + x^3 + x^3y$ & $y^2 + x^3y$ & \\
\hline 
$\llbracket12,4,3\rrbracket$& 6&1& 3 &  $1 + x + x^2 + x^3$ & $1 + x^2$ & \multirow{4}{*}{Square}\\ 
\cline{0-5} 
$\llbracket36,8,4\rrbracket$& 3&6& 3.55 & $1 + x + x^2y^3$ & $xy + xy^2 + xy^3$ & \\ 
\cline{0-5}
$\llbracket40,4,5\rrbracket$& 2&10& 2.5 & $x + y$ & $y + y^2 +y^3+y^4$ & \\ 
\cline{0-5}
$\llbracket70,10,5\rrbracket$& 5&7& 3.57 & $1 + y^2 + x+ xy^2$ & $xy + xy^2$ & \\ 
\hline 
\end{tabular}
\label{code_poly}
\end{center}
\end{table}
\endgroup

\subsection{Open Boundaries}
The highest code rates found for open-boundary codes are shown in Table \ref{OBC_code_distance}. Similar to the periodic-boundary codes, we scan different values of $l$ and $m$ ranging from 1 to 20, with the number of data qubits no more than $n_{max} = 200$, and an upper bound of distance is set to $d_{max} = 8$.

\begin{table}[ht]
\begin{center}
\begin{tabular}{ |c|c|c|c|c|c|}
\hline
 \multicolumn{6}{|c|}{Weight-4 Stabilizers}  \\ 
 \hline
Distance  & 3 & 4 & 5 & 6 & 7\\  
\hline
Square  & $\sfrac{1}{9}$ & $\sfrac{1}{16}$ & $\sfrac{1}{25}$ & $\sfrac{1}{36}$ & $\sfrac{1}{49}$\\
\hline
Hexagon  & $\sfrac{1}{9}$ & $\sfrac{1}{16}$ & $\sfrac{1}{25}$ & $\sfrac{1}{36}$ & $\sfrac{1}{49}$\\  
\hline
\hline
 \multicolumn{6}{|c|}{Weight-5 Stabilizers}  \\ 
 \hline
Distance  & 3 & 4 & 5 & 6 & 7\\  
\hline
Square  & $\sfrac{1}{6}$ & $\sfrac{1}{10}$ & $\sfrac{1}{18}$ & $\sfrac{1}{25}$ & $\sfrac{1}{32}$\\  
\hline
Hexagon  & $\sfrac{1}{6}$ & $\sfrac{2}{23}$ & $\sfrac{1}{18}$ & $\sfrac{1}{25}$ & $\sfrac{1}{34}$\\  
\hline
\hline
 \multicolumn{6}{|c|}{Weight-6 Stabilizers}  \\ 
 \hline
Distance  & 3 & 4 & 5 & 6 & 7\\  
\hline
Square  & $\sfrac{2}{9}$ & $\sfrac{1}{7}$ & $\sfrac{4}{45}$ & $\sfrac{4}{63}$ & $\sfrac{2}{43}$\\ 
\hline
Hexagon  & $\sfrac{1}{5}$ & $\sfrac{3}{29}$ & $\sfrac{3}{43}$ & $\sfrac{1}{20}$ & $\sfrac{3}{80}$\\ 
\hline
\end{tabular}
\caption{Highest code rates achieved for each code distance on different qubit connectivities under open boundary conditions after deep reduction.}
\label{OBC_code_distance}
\end{center}
\end{table}

\begin{figure}[ht]
\centering
\subfloat[]{\includegraphics[width = 0.45 \columnwidth]{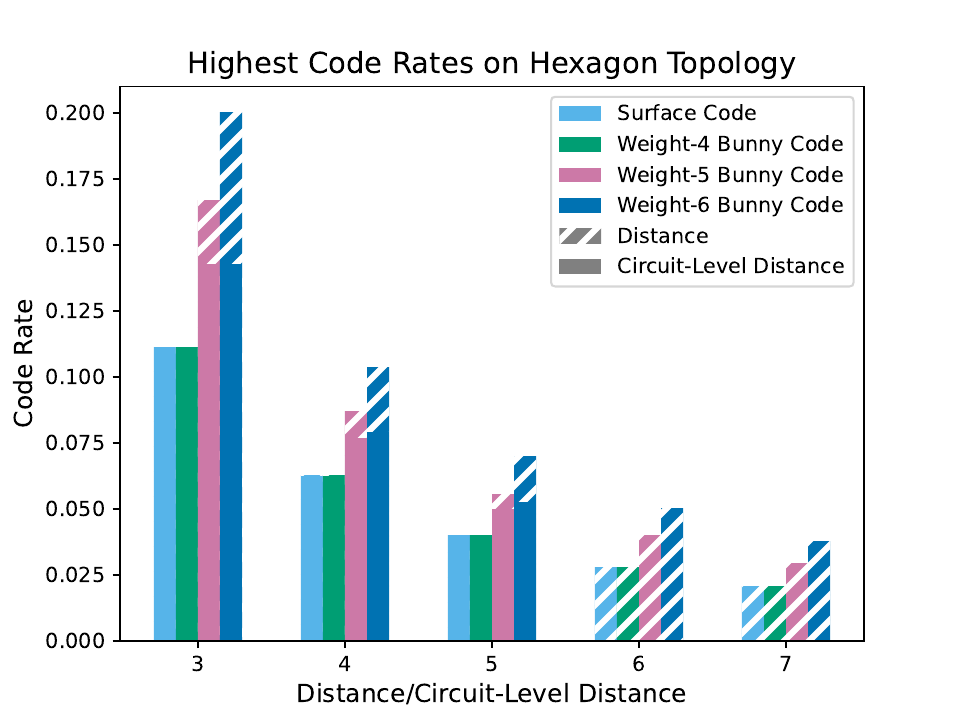}}\hspace{3mm}
\subfloat[]{\includegraphics[width = 0.45 \columnwidth]{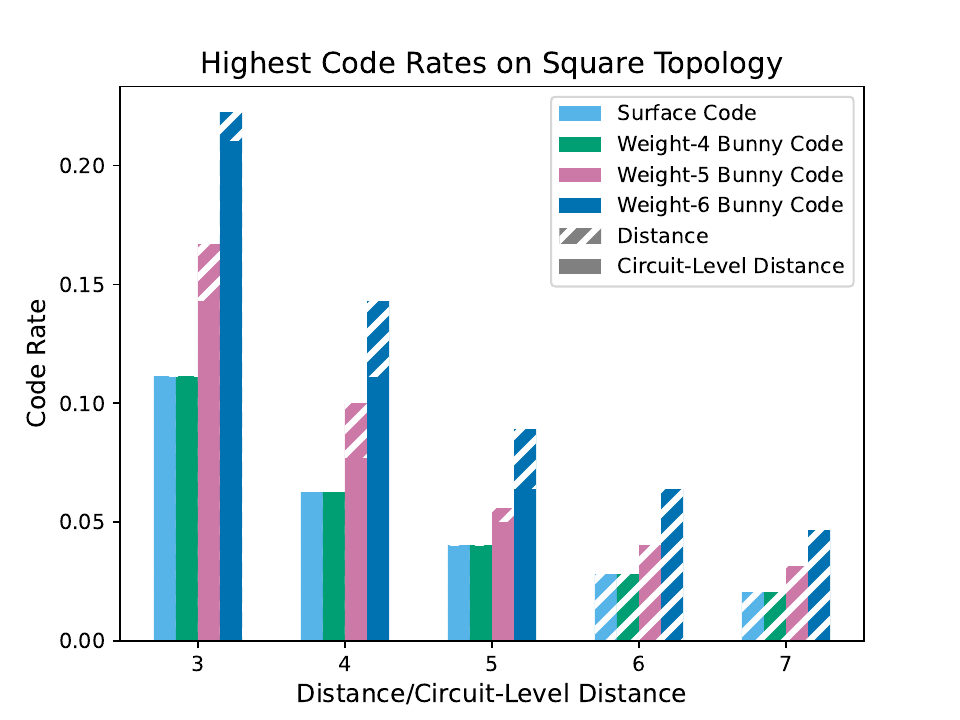}}\\
\caption{Highest code rates achieved for quantum error correction codes with different connectivities under open boundary conditions. (a) Hexagons. (b) Squares.}
\label{opc_rate}
\end{figure}

We see that Bunny codes with higher stabilizer weights still have an advantage in code rates compared to the surface code by a factor of ${\sim}2$, though this advantage is not as strong as in periodic boundaries. We stress again that those code rates are calculated from a nondeterministic heuristic of qubit reduction, thus only serve as a practical lower bound, and are not guaranteed to be the optimal code rate.

\begin{figure}[ht]
\centering
\includegraphics[width = 0.6 \columnwidth]{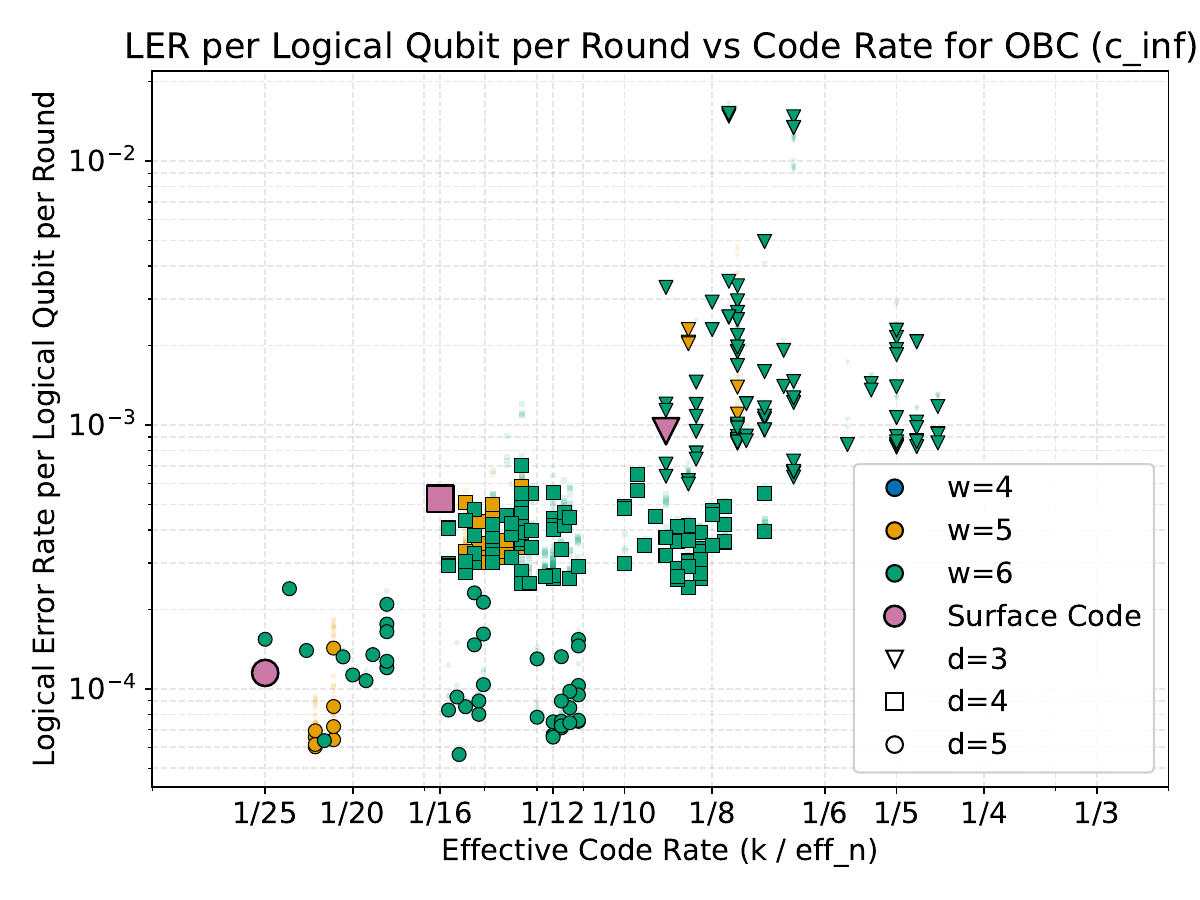}
\caption{The logical error rates per round per logical qubit of open-boundary Bunny codes and surface codes at a physical noise level of 0.001. For clarity of presentation, codes are grouped by their parameters stabilizers and $(n,k,d)$ and within each group only the code with the best logical performance is highlighted while the other codes are faded out. The colors of the markers correspond to the weights of their stabilizers, while the shapes correspond to their code distances. The performance of surface codes is highlighted in pink for reference.}
\label{obc_square_scatter}
\end{figure}

The logical error rates of some representative codes under open boundary conditions are shown in Fig.~\ref{ler_pbc}, and the larger scale evaluation is shown in Fig.~\ref{obc_square_scatter}. Bunny codes still exhibit significantly lower logical error rates than surface codes with comparable code rates, for example, in the $\llbracket 28,4,4\rrbracket$ code. It is worth noting that under open boundary conditions, we found that neither the code distance nor the circuit-level distance serves as a reliable predictor of logical performance under simulation, as circuits with different circuit-level distances can exhibit similar logical performance, while codes with the same code distance can have very different logical performance. Besides, during the deep reduction process, we only enforced the number of logical dimensions and the code distance to be the same. However, this process may effect on the circuit-level distance of the implementation of the code. Codes with higher code rate and good circuit level distance could be found through a more sophisticated qubit grafting procedure.

\begin{figure}[ht]
\centering
\subfloat[]{\includegraphics[width = 0.33 \columnwidth]{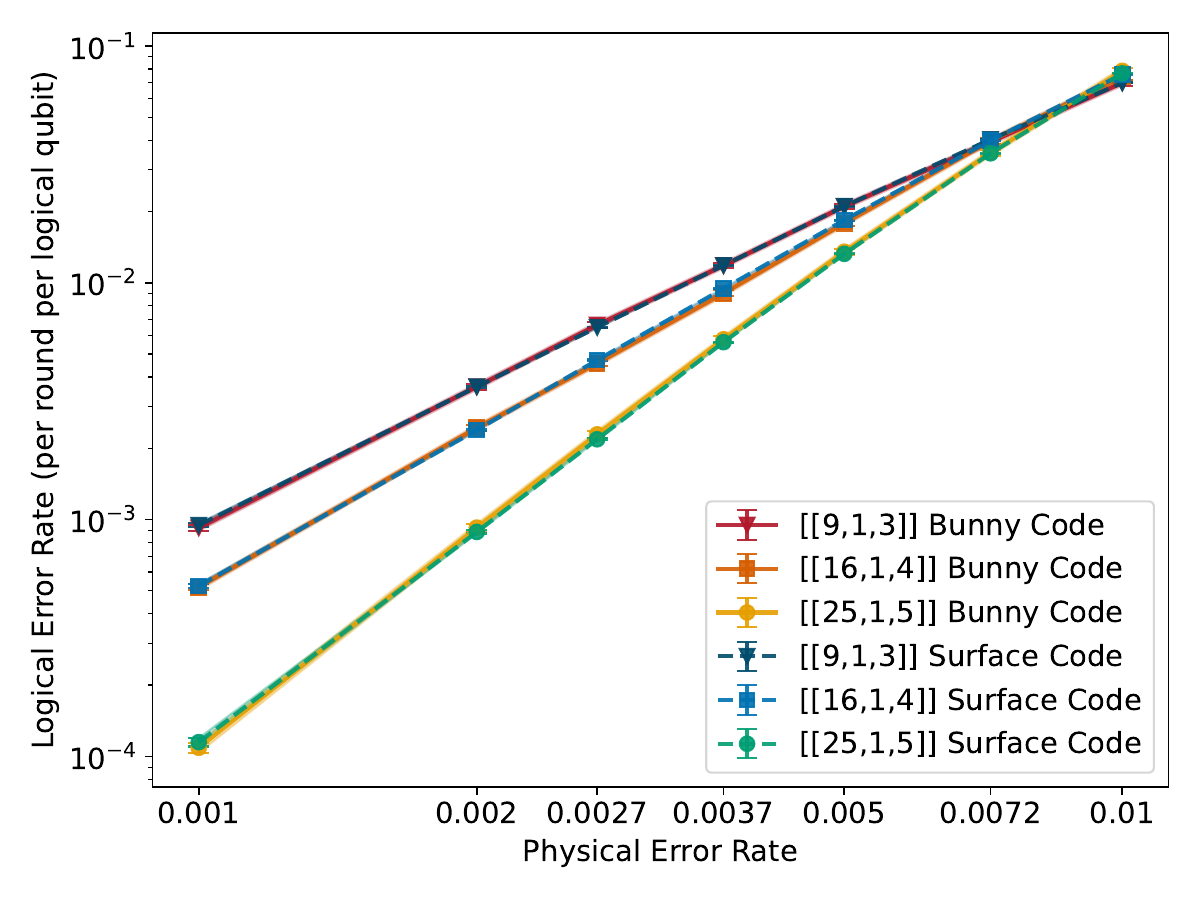}}
\subfloat[]{\includegraphics[width = 0.33 \columnwidth]{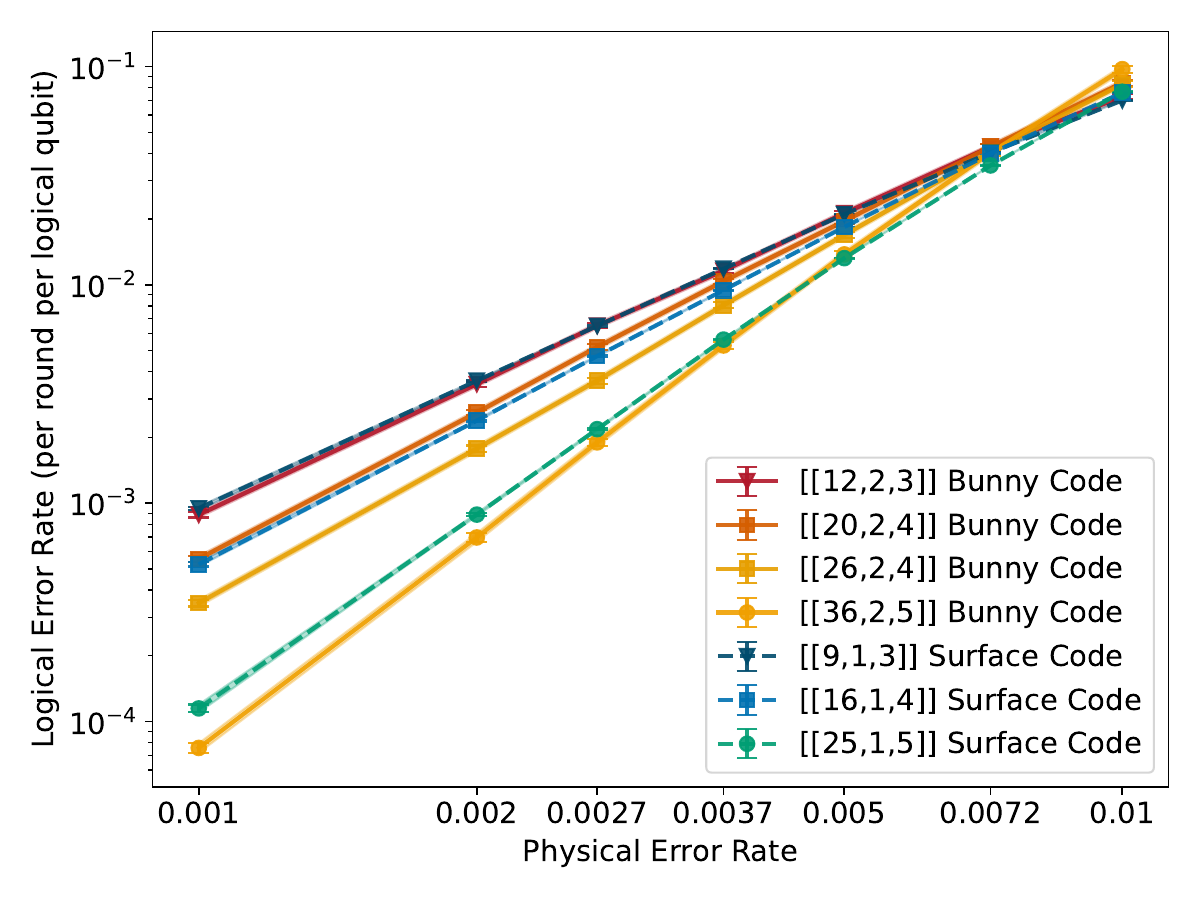}}
\subfloat[]{\includegraphics[width = 0.33 \columnwidth]{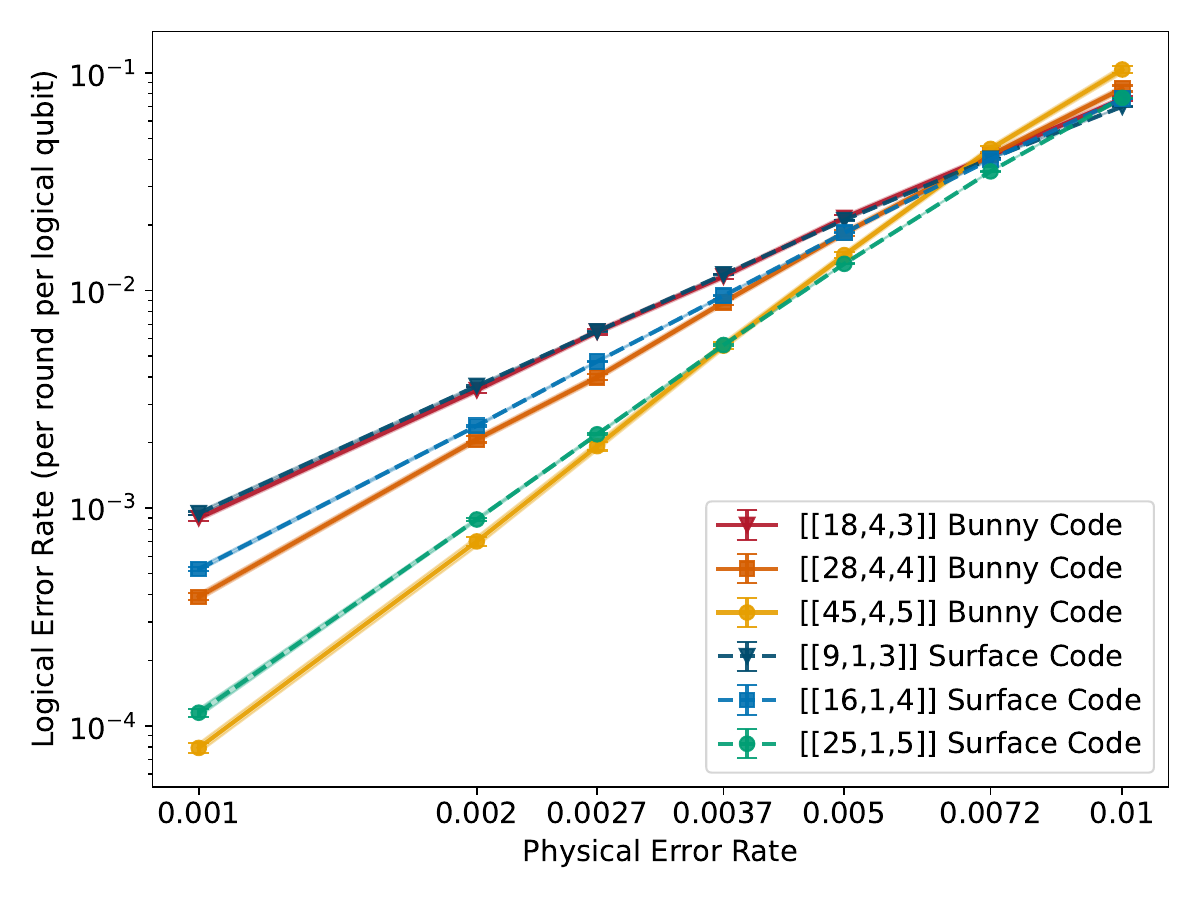}}\\
\caption{The logical error rate per round per logical qubit for some representative codes under open boundary conditions. (a) Weight-4 stabilizers. (b) Weight-5 stabilizers. (c) Weight-6 stabilizers.}
\label{ler_obc}
\end{figure}

Table \ref{obc_code_poly} includes the relevant parameters and the shapes of stabilizers in terms of their generating polynomials for some representative open-boundary codes. It is worth noting that during the nondeterministic deep reduction algorithm, entire columns or rows of qubits on the boundary of the code may be are reduced. Consequently, the same code could potentially be derived by reducing from smaller $l$ and $m$. Meanwhile, we also found that some of our codes are unique to open boundary conditions, in that they do not have a periodic-boundary counterpart - possibly a stabilized state without logical dimension. Our framework ensures that these open-boundary codes can be implemented on connectivity-constrained hardware. However, their theoretical origin could differ from those rigorously converted from periodic-boundary codes, thus exhibiting different properties.
\begingroup
\renewcommand{\arraystretch}{0.7}
\begin{table}[ht]
\small
\begin{center}
\caption{The parameters and generating polynomials of some of the open-boundary codes discussed in this paper. $n$ is the number of data qubits after deep reduction}
\label{code_parameter_obc}
\begin{tabular}{ |c|c|c|c|c|c|c| } 
\hline
$\llbracket n,k,d\rrbracket$&$l$&$m$&  $kd^2/n$ & $A$ & $B$ & Connectivity\\ 
\hline
\hline
$\llbracket12,2,3\rrbracket$& 4&3& 1.5 &\multirow{2}{*}{$y+xy+x^2$} & \multirow{2}{*}{$x+xy$} & \multirow{4}{*}{Square}\\ 
\cline{0-3}
$\llbracket36,2,5\rrbracket$& 8&5& 1.39 & & & \\ 
\cline{0-5}
$\llbracket20,2,4\rrbracket$& 5&10& 1.6 & $1+x+x^2y^2$ &  $xy + xy^2$ & \\ 
\cline{0-5}
$\llbracket26,2,4\rrbracket$& 7&5& 1.23 & $1 + x + x^2$  &  $xy + x$ & \\ 
\hline
\hline
$\llbracket18,4,3\rrbracket$& 4&5& 2 & $y^2 + xy^2 + x^2 $ & $x + xy + xy^2$ & \multirow{3}{*}{Square}\\ 
\cline{0-5}
$\llbracket28,4,4\rrbracket$& 7&7& 2.29 &  $1 + xy^2 +x^2y^2$ & $x + xy + xy^2$ & \\ 
\cline{0-5}
$\llbracket45,4,5\rrbracket$& 6&9& 2.22 & $1 + x +x^2y$ & $xy + xy^2 + xy^3$ & \\ 
\hline
\end{tabular}
\label{obc_code_poly}
\end{center}
\end{table}
\endgroup

\section{Discussion}
\label{sec:discussion}
In this work, we propose a general framework that, given the qubit connectivity and native two-qubit gate set of a hardware platform, exhaustively searches for translationally symmetric quantum low-density parity-check codes that can be efficiently implemented on that hardware. We find codes with weight-6 stabilizers that achieve the same code distance using only ${\sim}33\%$ and ${\sim}50\%$ physical qubits compared to the toric code and the surface code in periodic and open boundaries, respectively. Our results demonstrate that expanding the native two-qubit gate set enables superconducting hardware to efficiently perform syndrome extraction for qLDPC codes with non-local stabilizers. This offers a practical route toward the near-term experimental realization of high-performance qLDPC codes.

Notably, our assumption regarding the equivalent performance of the two types of two-qubit gates in our instruction set might not be entirely true in experiments. Certain hardware might be able to perform one type of two-qubit gate with significantly higher fidelity than the others, which will affect the resulting logical error rate. However, as we have previously mentioned, for a given set of stabilizers, we typically have multiple hardware-implementable syndrome extraction circuits for the stabilizers exhibiting similar performances under circuit-level simulations, as shown in Fig.~\ref{cld_ler_demo}. Thus, we can mitigate the aforementioned effect of asymmetric gate performance by choosing the syndrome extraction circuit with a greater proportion of the higher-fidelity two-qubit gate. Meanwhile, it is straightforward to see that the choice of using $\CNOT$ or $\CXSWAP$ gate in the first and final layers of the syndrome extraction circuit can be absorbed into the initial and final qubit configurations without affecting the main part of the circuit. Similarly, these gates can be replaced with the two-qubit gate favored by the hardware so that the experimental performances would be closer to our numerical simulations. In addition, as we have seen in the results section, certain Bunny codes outperform the toric code with a similar code rate even when the physical error rate is doubled. This advantage in performance persists even against the highly pessimistic reference scenario where every $\CXSWAP$ gate is decomposed into two $\CNOT$ gates, particularly when combined with the error mitigation strategy introduced earlier.

Furthermore, though we assume that our hardware supports two native two qubit gates during our search, some of our results require only $\CXSWAP$ gates, such as the $\llbracket12,4,3\rrbracket$, $\llbracket18,6,3\rrbracket$, and $\llbracket30,6,4\rrbracket$ codes generated by the action sequence in Table \ref{ex_actions}. As we previously discussed, we may alter the gate types in the first and last layers by adjusting the initial and final qubit configurations, thereby converting all the gates in Table \ref{ex_actions} into $\CXSWAP$ gates. Codes that only employ $\CXSWAP$ gates demand less calibration overhead on the quantum hardware, and we leave a systematic search of those $\CXSWAP$-only codes for future work.

There are many extensions that could stem from this work. First, though our framework allows us to evaluate arbitrary hardware connectivity satisfying the constraint of translational symmetry, we have confined our search to some of the most basic topologies. In a more thorough study, we could combine the current approach with long-range couplers to examine whether better codes can be implemented on more sophisticated hardware. Furthermore, we can search for the favorable connectivities that support high-rate quantum error correction codes while satisfying high-level hardware constraints, such as the qubit degree, the Tanner graph thickness, the average and longest coupler lengths, etc. In terms of the native two-qubit gate set, our search incorporated only $\CNOT$ and $\CXSWAP$ as our native two-qubit gates. Additional routing flexibility could be obtained by including $\SWAP$ gates as well, at the cost of tolerating some noise overhead that can be regulated by limiting the length of the action sequence. Although $\SWAP$ gates are not commonly available as native two-qubit gates and would likely require longer gate times than the other gates, thereby complicating the gate-layer scheduling, the existing gate scheme could nevertheless enable them to facilitate longer-range interactions between ancilla and data qubits. More technically, by improving the performance of the many algorithms used in our analysis, we could significantly broaden the scope of our search space in terms of code distance, stabilizer weights, and potentially identifying better open boundary codes.

\appendix

\section{A List of Sequence of Actions}
In Table \ref{code_ninst} we list the sequences of actions that generate the circuits we used for simulation in the Results Section. The actions are displayed in terms of numerical index, with the correspondence described in Table \ref{ninst_dict}. As we stated previously, it is often the case that each set of stabilizers can be measured by more than one syndrome extraction circuit, so here we present only the circuits with the lowest logical error rate in simulation for each set of $n,k,d$.

\begin{table}[ht]
\small
\begin{center}
\caption{The parameters and sequence of actions of some of the codes simulated in the result section of this paper.}
\begin{tabular}{ |c|c|c|c|} 
\hline
$\llbracket n,k,d\rrbracket$&l&m&  Sequence of Actions\\ 
\hline
\hline
$\llbracket12,2,3\rrbracket$& 2&3& 0, 12, 4, 6, 11, 15, 2, 8\\ 
\hline
$\llbracket16,2,4\rrbracket$& 2&4& 0, 12, 4, 6, 15, 11, 2, 8 \\ 
\hline
$\llbracket30,2,5\rrbracket$& 3&5& 0, 12, 4, 7, 11, 14, 0, 12\\ 
\hline
$\llbracket10,2,3\rrbracket$& 1&5& 0, 12, 5, 7, 15, 11, 2, 8\\ 
\cline{0-2}
$\llbracket28,2,5\rrbracket$& 2&7& 0, 12, 5, 7, 15, 11, 2, 8\\ 
\hline
\hline
$\llbracket18,4,3\rrbracket$& 3&3& 1, 3, 8, 5, 7, 14, 10, 0, 13, 8\\ 
\hline
$\llbracket24,4,4\rrbracket$& 3&4& 1, 3, 8, 4, 6, 15, 11, 1, 13, 8\\ 
$\llbracket30,4,5\rrbracket$& 3&5& 0, 3, 9, 5, 7, 15, 11, 1, 13, 8\\ 
$\llbracket54,4,5\rrbracket$& 3&9& 0, 12, 5, 7, 5, 10, 15, 10, 0, 12\\ 
\hline
\hline
$\llbracket12,4,3\rrbracket$& 1&6&0, 8, 4, 7, 5, 6, 14, 11, 15, 10, 0, 8\\ 
\hline
$\llbracket16,4,4\rrbracket$& 4&2&0, 5, 15, 3, 1, 3, 9, 13, 9, 5, 15, 12\\ 
\hline
$\llbracket36,8,4\rrbracket$& 3&6&0, 3, 9, 5, 7, 5, 15, 11, 15, 3, 0, 8\\ 
\hline
$\llbracket40,4,5\rrbracket$& 2&10&0, 12, 4, 7, 5, 6, 15, 11, 15, 11, 2, 8\\ 
\hline
$\llbracket70,10,5\rrbracket$& 5&7&1, 3, 9, 13, 5, 7, 15, 11, 3, 0, 13, 8\\ 
\hline
\end{tabular}
\label{code_ninst}
\end{center}
\end{table}

\begin{table}[ht]
\small
\begin{center}
\caption{The index of all the action on the square grid connectivity.}
\begin{tabular}{ |c|c|c| } 
\hline
Index & Coupler & Gate\\ 
\hline
0 & $(0,0) \leftrightarrow (1,0)$ & $\CNOT$ \\
\hline 
1 & $(0,0) \leftrightarrow (1,0)$ & $\CXSWAP$ \\
\hline 
2 & $(0,0) \leftrightarrow (-1,0)$ & $\CNOT$ \\
\hline 
3 & $(0,0) \leftrightarrow (-1,0)$ & $\CXSWAP$ \\
\hline 
4 & $(0,0) \leftrightarrow (0,1)$ & $\CNOT$ \\
\hline 
5 & $(0,0) \leftrightarrow (0,1)$ & $\CXSWAP$ \\
\hline 
6 & $(0,0) \leftrightarrow (0,-1)$ & $\CNOT$ \\
\hline 
7 & $(0,0) \leftrightarrow (0,-1)$ & $\CXSWAP$ \\
\hline 
8 & $(0,1) \leftrightarrow (1,1)$ & $\CNOT$ \\
\hline 
9 & $(0,1) \leftrightarrow (1,1)$ & $\CXSWAP$ \\
\hline 
10 & $(1,0) \leftrightarrow (1,1)$ & $\CNOT$ \\
\hline 
11 & $(1,0) \leftrightarrow (1,1)$ & $\CXSWAP$ \\
\hline 
12 & $(0,1) \leftrightarrow (-1,1)$ & $\CNOT$ \\
\hline 
13 & $(0,1) \leftrightarrow (-1,1)$ & $\CXSWAP$ \\
\hline 
14 & $(1,0) \leftrightarrow (1,-1)$ & $\CNOT$ \\
\hline 
15 & $(1,0) \leftrightarrow (1,-1)$ & $\CXSWAP$ \\
\hline 
\end{tabular}
\label{ninst_dict}
\end{center}
\end{table}

\section*{Acknowledgement}
Part of this work was carried out during the internship of R. Z. and X. Y. at TraverseQuantum Inc. This work is supported by Zhongguancun Laboratory and the National Key Research and Development Program of China (Grant No.2025YFE0200900).

\bibliographystyle{quantum}
\bibliography{Coupler_Code}

\end{document}